%% file: main.tex
\documentclass[a4paper,fleqn]{cas-dc}
\usepackage{amsmath,amsfonts}
\usepackage{algorithmic}
\usepackage{array}
\usepackage[caption=false,font=normalsize,labelfont=sf,textfont=sf]{subfig}
\usepackage{textcomp}
\usepackage{stfloats}
\usepackage{url}
\usepackage{verbatim}
\usepackage{graphicx}
\def\BibTeX{{\rm B\kern-.05em{\sc i\kern-.025em b}\kern-.08em
    T\kern-.1667em\lower.7ex\hbox{E}\kern-.125emX}}
\usepackage{balance}
\usepackage{etoolbox}
\newenvironment{cequation}{
	\makeatletter
	\setbool{@fleqn}{false}
	\makeatother
	\begin{equation}
	}{\end{equation}}

\usepackage[labelsep=period]{caption}

\input{cmd_def}

\begin{document}
\let\WriteBookmarks\relax
\let\printorcid\relax 
\def\floatpagepagefraction{1}
\def\textpagefraction{.001}

\title[mode = title]{Efficient Reduction of Interconnected Subsystem Models using Abstracted Environments}

\shorttitle{Efficient Reduction of Interconnected Subsystem Models using Abstracted Environments}
\shortauthors{L. Poort et~al.}

\tnotemark[1] 
\tnotetext[1]{This work was supported by Holland High Tech \textbar \ TKI HSTM via the PPS allowance scheme for public-private partnerships and ASML.}


\author[1]{{Luuk Poort}}
\cormark[1]
\cortext[cor1]{Corresponding author \\ E-mail address: \href{l.poort@tue.nl}{l.poort@tue.nl}}
\credit{Conceptualization, Methodology, Formal analysis, Software, Writing - original draft, Writing - review \& editing}

\author[2]{{Bart Besselink}}
\credit{Supervision, Conceptualization, Methodology, Writing - review \& editing}
\author[1]{{Rob H.B. Fey}}
\credit{Supervision, Conceptualization, Methodology, Writing - review \& editing}
\author[1]{{Nathan van de Wouw}}
\credit{Supervision, Conceptualization, Methodology, Writing - review \& editing}

\affiliation[1]{organization={Department of Mechanical Engineering, Eindhoven University of Technology},
                postcodesep={}, 
                postcode={5600 MB}, 
                city={Eindhoven},
                country={The Netherlands}}

\affiliation[2]{organization={Bernoulli Institute for Mathematics, Computer Science and Artificial Intelligence, University of Groningen},
                postcodesep={}, 
                postcode={9700 AB}, 
                city={Groningen},
                country={The Netherlands}}

\maketitle

%
%
\begin{abstract}
    We present two frameworks for structure-preserving model order reduction of interconnected subsystems, improving tractability of the reduction methods while ensuring stability and accuracy bounds of the reduced interconnected model. Instead of reducing each subsystem independently, we take a low-order abstraction of its environment into account to better capture the dynamics relevant to the external input-output behaviour of the interconnected system, thereby increasing accuracy of the reduced interconnected model. This approach significantly reduces the computational costs of reduction by abstracting instead of fully retaining the environment. The two frameworks differ in how they generate these abstracted environments: one abstracts the environment as a whole, whereas the other abstracts each individual subsystem. By relating \emph{low-level} errors introduced by reduction and abstraction to the resulting \emph{high-level} error on the interconnected system, we are able to translate high-level accuracy requirements (on the reduced interconnected system) to low-level specifications (on abstraction and reduction errors) using techniques from robust performance analysis. By adhering to these low-level specifications, restricting the introduced low-level errors, both frameworks automatically guarantee the accuracy and stability of the reduced interconnected system. We demonstrate the effectiveness of both frameworks by applying them to a structural dynamics model of a two-stroke wafer stage, achieving improved accuracy and/or greater reduction compared to an existing method from literature.
\end{abstract}

%
%

\section{Introduction} \label{sec:intro}
Complex dynamical systems are typically composed of (many) interconnected subsystems, which are designed and modeled by separate engineering teams. Examples include large machine assembly models, models of power grids and complex control systems. Even when individual subsystem models are of moderate order, the model of the interconnected system, which consists of the assembly of all subsystem models, quickly becomes of such high order that dynamical analysis and controller design are computationally infeasible. To overcome these computational limitations, model order reduction methods are employed to approximate the high-order model by a surrogate model, which is typically of considerably lower order.

\begin{figure}\hspace{1mm}
    \subfloat[\small]{%
      \includegraphics[width=.21\linewidth]{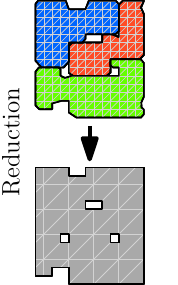} \label{fig:cpldRed}
    }\hfill
    \subfloat[\small]{%
      \includegraphics[width=.5\linewidth]{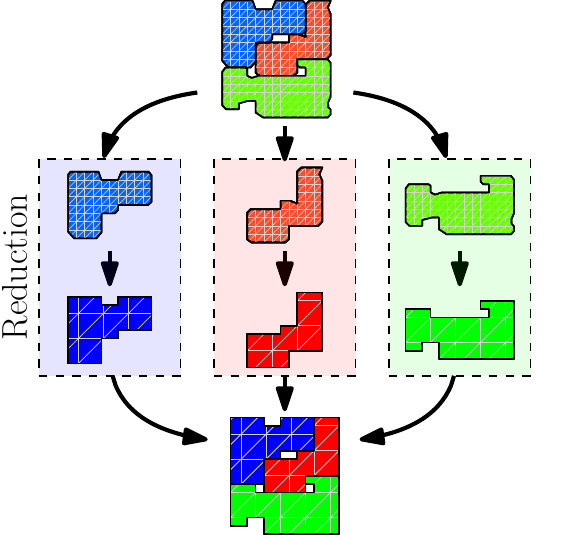} \label{fig:indepSSRed}
    }\hfill
    \subfloat[\small]{%
      \includegraphics[width=.21\linewidth]{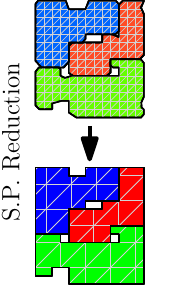} \label{fig:SPRed}
    }\hspace{1mm}
    \caption{Approaches to interconnected model reduction: (a) closed-loop reduction of the interconnected model, (b) open-loop reduction of independent subsystem models and (c) structure-preserving reduction.}
    \label{fig:interc.sysred}
\end{figure}

When reducing a model of interconnected subsystems, there exists three main approaches \cite{Lutowska2012ModelApproximations,Reis2008ASystems,Benner2015ModelSurvey,Sandberg2009}, as visualized in \autoref{fig:interc.sysred}. First of all, one can reduce the full, interconnected model directly, i.e., via \emph{closed-loop} reduction (\autoref{fig:cpldRed}), which generally results in a highly accurate reduced-order model \cite{DeKlerk2008GeneralTechniques,Besselink2013AControl,Antoulas2005ApproximationSystems}. However, as the interconnected system is of very high order, the reduction approach itself can become computationally infeasible. In addition, this approach does not retain the essential structure of the interconnected system, such that it is no longer possible to discern individual reduced subsystem models \cite{Sandberg2009,Vandendorpe2008ModelSystems}. 

To preserve the interconnection structure, subsystem models can also be reduced independently, i.e., via \emph{open-loop} reduction, such that the reduced, interconnected model consists of an interconnection of reduced subsystem models (\autoref{fig:indepSSRed}). This modular approach aligns with the modular design of the subsystems, providing each design team with greater flexibility and preserving the essential structure of the interconnected system \cite{Reis2008ASystems,Janssen2024ModularPerspective}. However, independent reduction of the subsystem models may not retain the dynamics that is most important for the interconnected system, such that the accuracy of the reduced interconnected system may be low \cite{Sandberg2009}.

Lastly, it is possible to reduce the individual subsystem models based on the interconnected system dynamics, i.e., via \emph{structure-preserving} reduction (\autoref{fig:SPRed}), thereby retaining the interconnection structure \cite{Sandberg2009,Vandendorpe2008ModelSystems,Wortelboer1994FrequencyConfiguration,Poort2024Balancing-BasedSystems,Kessels2022Sensitivity-BasedReduction,Kim2018ADisplacement,Li2005Structure-PreservingFormulation,Li2005StructuredLMIs}. With respect to open-loop reduction, this modular, structure-preserving approach can significantly improve the accuracy of the reduced interconnected model. Unfortunately, the evaluation of the interconnected dynamics comes with a high computational cost, similar to closed-loop reduction. Due to these computational limitations, open-loop reduction remains the most popular method for reducing large interconnections of high-order models in practice.

In our previous paper \cite{Poort2024AbstractedReduction}, we have therefore suggested an alternative point of view. In the structure-preserving reduction of a single subsystem, its environment, comprising all remaining subsystems, can be regarded as a weighting function used to identify the essential subsystem dynamics to retain \cite{Anderson1987ControllerApproaches}. Because this weighting function is often of excessively high order, we propose in \cite{Poort2024AbstractedReduction} to instead use a low-order \emph{abstraction} of the environment model as a weighting function. 

The use of such abstractions forms the basis for the framework of abstracted reduction, as proposed in \cite{Poort2024AbstractedReduction}, to improve the tractability of structure-preserving reduction methods. This framework involves an initial open-loop abstraction of the environment, followed by a structure-preserving reduction of the one subsystem only, while connected to this abstracted environment. Using tools from robust performance, a priori requirements can be posed on the accuracy of both the environment's abstraction and the subsystem's structure-preserving reduction that guarantee the stability and accuracy of the reduced interconnected system model.

However, the abstracted reduction framework of \cite{Poort2024AbstractedReduction}, while representing an essential stepping stone to the work in the current paper, only allows accuracy and stability guarantees for the reduction of a \emph{single system} in connection to an environment model, but not for the reduction of general interconnected systems. The current paper therefore extends the methods of \cite{Poort2024AbstractedReduction} to widen the applicability of the abstracted reduction framework to the full generality of the reduction of all subsystems of an interconnected system.

As our first main contribution, we present two variants of the framework of \emph{abstracted} reduction of interconnected subsystem models. In both variants, each subsystem is reduced while connected to an abstraction of its environment model. However, in the first variant, the environment of each subsystem is first composed of all other subsystems and subsequently reduced to a low-order abstraction, whereas in the second variant, the abstracted environment of each subsystem is composed by interconnecting low-order abstracted subsystem models for the other subsystems.

Secondly, we employ techniques from robust performance analysis to quantitatively assess how relying solely on an environment abstraction influences the resulting accuracy of the reduced interconnected system model. This allows us to establish a priori requirements on the accuracy of both the abstraction and the structure-preserving reduction, based on user-defined specifications on the reduced interconnected system model's accuracy. These accuracy requirements are then leveraged to formulate a systematic approach for the abstracted reduction framework. This approach automatically generates reduced-order subsystem models which guarantee the stability and the satisfaction of the accuracy specification of the reduced, interconnected system model.

Finally, both variants of the abstracted reduction framework are evaluated using a structural dynamics model, consisting of three interconnected subsystem models. Compared to the subsystem reduction approach of \cite{Janssen2023ModularApproach}, our methods demonstrate clear advantages by incorporating the environment into the reduction process.

The remainder of this paper is structured as follows. Firstly, in \autoref{sec:prob}, the system representation is introduced and the problem is formally defined. Then, in \autoref{sec:absred}, two generalized abstracted reduction frameworks are introduced to improve the computational efficiency for the structure-preserving reduction of interconnected systems. To also ensure the accuracy and stability of the reduced, interconnected system, the corresponding error dynamics are evaluated in \autoref{sec:err_analysis}. To avoid determining the reduction orders by trial and error, a systematic approach to both abstracted reduction frameworks is subsequently introduced in \autoref{sec:spec}. The proposed methods are finally evaluated in \autoref{sec:case_study_red} by means of a structural dynamics benchmark model after which conclusions are drawn in \autoref{sec:con}.

\emph{Notation:}
In this paper, sets are generally indicated by blackboard-bold symbols, such as $\R$, $\R_{>0}$ and $\C$, which denote the set of real, positive real and complex numbers, respectively. $\R^{m\times p}$ and $\C^{m\times p}$ indicate matrices of real and complex numbers with $m$ rows and $p$ columns. Given a complex matrix $A$, $A^\top$ and $A^H$ denote its transpose and conjugate transpose, respectively, $\|A\|$ denotes its spectral norm, $A\succ 0$ and $A\succeq 0$ denote that $A$ is positive definite and positive semi-definite, respectively, and $A = \diag(A_1,A_2)$ denotes a block-diagonal matrix of submatrices $A_1$ and $A_2$. The zero and identity matrices are denoted by $O$ and $I$, respectively, while $I_n$ denotes an identity matrix of size $n\times n$. A transfer function matrix is denoted as $\sys(s)$, with $s$ the Laplace variable and $\|\sys\|_\infty$ its $\Hnrm{\infty}$-norm. The set of all proper, real rational stable transfer matrices is denoted by \!$\RH$. 

%
%
\section{Problem setting}\label{sec:prob}

\subsection{Interconnected systems}\label{ssec:sys_repr}
\begin{figure}
    \subfloat[\small\label{fig:coupling_diag}]{%
      \includegraphics[width=0.45 \linewidth]{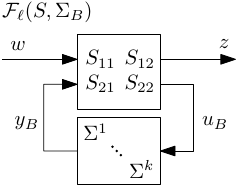}
    }
    \hfill
    \subfloat[\small \label{fig:coupling_diag_red}]{%
      \includegraphics[width=0.45\linewidth]{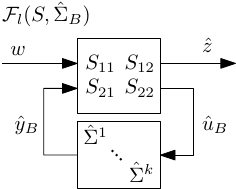}
    }
    \caption{(a) Lower LFT of $\S(s)$ and $\sys_B(s)$, constituting the interconnected model $\llft(\S,\sys_B)$ and (b) lower LFT of $\S(s)$ and $\sysr_B(s)$, constituting the reduced, interconnected model $\llft(\S,\sysr_B)$.}
    \label{fig:coupling_diags}
\end{figure}

Consider $k$ subsystems and their interconnection, modeled by the proper, real rational transfer function matrices $\ssys{1}(s),\dots,\ssys{k}(s)$ and $\S(s)$, respectively, as visualized in \autoref{fig:coupling_diag}. Each subsystem model $\ssys{j}(s)$ has inputs $u^{j}$ of dimension $m_j$, outputs $y^{j}$ of dimension $p_j$ and McMillan degree (order) $n_{\sys,j}$, for $j\in \{1,\dots,k\}$. The interconnection dynamics model $\S(s)$ has inputs $\left[\begin{smallmatrix} w \\ y_B \end{smallmatrix}\right]$ of dimension $m_C+p_B$, and outputs $\left[\begin{smallmatrix} z \\ u_B \end{smallmatrix}\right]$ of dimension $p_C+m_B$ and order $n_S$. 

To formally define the interconnected system, we first define the parallel interconnection of all subsystems as $\sys_B \coloneqq \diag(\ssys{1},\dots,\ssys{k})$ with inputs and outputs, respectively, as
\beq \label{eq:sysb_IOs}
u_B = \begin{bmatrix}    u^{1} \\ \vdots \\ u^{k} \end{bmatrix}, \qquad 
y_B = \begin{bmatrix}    y^{1} \\ \vdots \\ y^{k} \end{bmatrix},
\eeq
such that $m_B = m_1 + \dots + m_k$ and $p_B = p_1 + \dots + p_k$.

The interconnected system is subsequently composed of $\S(s)$ and $\sys_B(s)$, by means of a lower linear fractional transformation (LFT), as defined next (see also \cite[Def.~9.2]{Zhou1998EssentialsControl} and \cite[Lem.~5.1]{Zhou1998EssentialsControl}).
\begin{defn} \label{def:llft_ulft_wellposed}
    Let $P(s) = \left[\begin{smallmatrix}  P_{11}(s) & P_{12}(s)\\ P_{21}(s) & P_{22}(s) \end{smallmatrix}\right]$, $M_l(s)$ and $M_u(s)$ be proper, real rational transfer function matrices of dimensions $(p_1+p_2)\times(m_1+m_2)$, $m_2\times p_2$ and $m_1\times p_1$, respectively. Then, we define the lower and upper LFTs as 
    \begin{align}
        \llft(P,M_l) &= P_{12}M_l(I-P_{22}M_l)^{-1}P_{21}+P_{11}, \label{eq:llft_def} \\ 
        \ulft(P,M_u) &= P_{21}M_u(I-P_{11}M_u)^{-1}P_{12}+P_{22},\label{eq:ulft_def}
    \end{align}    
    which are said to be \emph{well-posed} if $I-P_{22}M_l$ and $I-P_{11}M_u$ have a proper real rational inverse, respectively.
\end{defn}

Using this definition, the interconnection of all subsystem models $\ssys{j}$, gathered in $\sys_B$, and the interconnection dynamics $S$, as in \autoref{fig:coupling_diag}, can be written as $\llft(S,\sys_B)$. This interconnected system model $\llft(S,\sys_B)$ has external inputs $w$, external outputs $z$ and order $n_C \leq n_B + n_S$.

\begin{figure}
    \subfloat[\small\label{fig:coupling_diag_env}]{%
      \includegraphics[width=0.45 \linewidth]{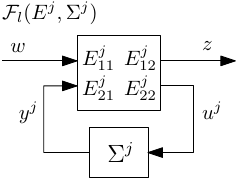}
    }
    \hfill
    \subfloat[\small \label{fig:env_expr}]{%
      \includegraphics[width=0.45\linewidth]{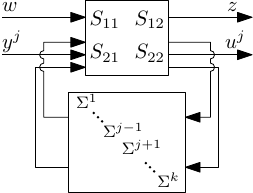}
    }
    \caption{(a) Lower LFT of $\senv{j}$ and $\ssys{j}$, constituting the interconnected model $\llft(\senv{j},\ssys{j}) = \llft(\S,\sys_B)$ and (b) lower LFT of $\S$ and all $\ssys{l}$ for $l \in \{1,\dots,k\}\backslash \{j\}$, constituting $\senv{j}$.}
    \label{fig:env_coupling_diags}
\end{figure}

It will turn out to be useful to express the interconnected model $\llft(\S,\sys_B)$ in an alternative (but equivalent) manner. To do so, we write $\llft(\S,\sys_B)$ as the interconnection of a single isolated subsystem $\ssys{j}$ and its environment $\senv{j}$, denoted as $\llft(\senv{j},\ssys{j})$ and visualized in \autoref{fig:coupling_diag_env}. This environment $\senv{j}$ is defined as the interconnection between $\S$ and all remaining subsystems $\ssys{l}$ for $l \in \{1,\dots,k\}\backslash \{j\}$, as shown in \autoref{fig:env_expr}, such that $\senv{j}$ has inputs $\left[\begin{smallmatrix} w \\ y^{j} \end{smallmatrix}\right]$ and outputs $\left[\begin{smallmatrix} z \\ u^{j} \end{smallmatrix}\right]$.

\begin{defn}\label{def:senv}
    Given $\S(s)$ and $\ssys{1}(s),\ \dots,\ \ssys{k}(s)$, the environment $\senv{j}(s)$ of each $\ssys{j}(s)$ is given as
    \begin{equation}
        \senv{j} \coloneqq \llft\Big(\Sb{j},\ssysb{j}_B\Big),
    \end{equation}
    where $\ssysb{j}_B$ is the parallel connection of all subsystems excluding $\ssys{j}$, such that
    \begin{equation}\label{eq:def_yb_ub_sysb}
        \ssysb{j}_B = \diag\!\big(\ssys{1},\dots,\ssys{j-1},\ssys{j+1},\dots,\ssys{k}\big).
    \end{equation}
    Moreover, $\Sb{j}$ is obtained by permuting the rows and columns of $S$ such that
    \begin{equation}
        \mbox{\small $\left[\begin{array}{c}            
            z \\ u^{j} \\\hline u^1\hspace{-3mm}\\\vdots\\u^{j-1}\\ u^{j+1}\\\vdots\\u^k
        \end{array}\right]$} =
        \Sb{j}\,
        \mbox{\small $\left[\begin{array}{c}
            w \\ y^{j} \\\hline y^1\\\vdots\\y^{j-1}\\y^{j+1}\\\vdots\\y^k
        \end{array}\right]$}.
    \end{equation}
    Specifically, $\Sb{j}$ is given as
    \begin{equation}\label{eq:def_Sb}
        \Sb{j} = \left[\begin{array}{cc|c}
            \S_{11} & \S_{12}^{j} & \S_{12}^{l}\\
            \S_{21}^{j} & \S_{22}^{j,j} & \S_{22}^{j,l}\\ \hline
            \S_{21}^{l} & \S_{22}^{l,j} & \S_{22}^{l,l}
        \end{array}\right],
    \end{equation}
    where superscripts $j$ and $l$ indicate the partitions related to $u^j$ or $y^j$ and $u^l$ or $y^l$, for $l \in \{1,\dots,k\}\backslash \{j\}$, respectively.
\end{defn}

We emphasize that, for all $j$,
\begin{equation}
    \llft(S,\sys_B) = \llft(\senv{j},\ssys{j})
\end{equation}
by \autoref{def:senv}.

We adopt the following assumption to ensure that both the environment models and the interconnected system are well-defined and internally stable, using \cite[Corollary~5.2]{Zhou1998EssentialsControl}.
\begin{ass}\label{ass1} 
    We have $\ssys{1},\dots,\ssys{k}\in\RH$ and $\S \in\RH$ and the environments $\senv{j}$ and interconnected system $\llft(\S,\sys_B) = \llft(\senv{j},\ssys{j})$ for $j = 1,\dots,k$ are well-posed and internally stable, i.e., $\ssysb{j}_B(I-\S_{22}^{l,l}\ssysb{j}_B)^{-1} \in \RH$ and $\sys_B(I-S_{22}\sys_B)^{-1} \in \RH$. Particularly, we have $\senv{1},\,\dots,\,\senv{k},\, \llft(\S,\sys_B)\in \RH$.
\end{ass}

\subsection{Structure-preserving model reduction} \label{ssec:structpres}
The objective of structure-preserving model reduction is to find transfer function matrices $\ssysr{j}(s)$ of order $r_{\sys,j} < n_{\sys,j}$ for all $j = 1,\dots,k$, such that $\llft(\S,\sysr_B)$, as depicted in \autoref{fig:coupling_diag_red}, is well-posed, stable and accurately approximates $\llft(\S,\sys_B)$ in terms of input-output behaviour, where $\sysr_B \coloneqq \diag(\ssysr{1},\dots,\ssysr{k})$. In other words, we reduce each subsystem model such that the interconnected system reduction error
\begin{equation}\label{eq:ec_init_def}
    \ec \coloneqq \llft(\S,\sysr_B)- \llft(\S,\sys_B)
\end{equation} 
is small in a suitable sense. 

Structure-preserving reduction methods such as those in \cite{Sandberg2009,Vandendorpe2008ModelSystems,Poort2024Balancing-BasedSystems,Cheng2019BalancedSystems} aim to find an accurate reduced-order model $\llft(\S,\sysr_B)$, but are computationally costly such that they are only applicable to interconnected systems of limited order $n_C$. In addition, most methods either do not guarantee stability of $\llft(\S,\sysr_B)$, e.g. \cite{Sandberg2009,Vandendorpe2008ModelSystems}, or require additional system properties such as passivity \cite{Poort2024Balancing-BasedSystems,Cheng2019BalancedSystems}.

A further, intrinsic limitation of structure-preserving reduction methods is the need for \emph{all} subsystem models to reduce a single one. In a modular design process, where subsystems are designed in parallel, all other subsystems are generally not (yet) available or only a rough estimate is available.

\subsection{Abstracted reduction} \label{ssec:absred_prelim}
In our previous work \cite{Poort2024AbstractedReduction}, we presented the framework of abstracted reduction to improve the tractability of structure-preserving reduction methods for the reduction of a single system $\sys(s)$ within an environment $\env(s)$, constituting the interconnected model $\llft(\env,\sys)$. Instead of using a structure-preserving reduction method to reduce $\sys$ within $\llft(\env,\sys)$ directly, we first reduced the environment $\env(s)$ to an \emph{abstraction} $\enva(s)$ and subsequently reduce the system $\sys(s)$ within $\llft(\enva,\sys)$ to $\sysr(s)$. By first abstracting $\env(s)$ to $\enva(s)$ in open loop with a cheap reduction method, the computational cost of the structure-preserving reduction of $\llft(\enva,\sys)$ can be significantly reduced.

Subsequently, tools from robust performance analysis were used to relate prescribed specifications on $\ec$, as defined in \autoref{eq:ec_init_def}, to requirements on the accuracy on $\enva$ and on $\llft(\enva,\sysr)$. Any approximation satisfying these requirements for the abstraction (from $\env$ to $\enva)$ and reduction (from $\sys$ to $\sysr$) then guarantees $\llft(\env,\sysr)$ to be stable and to meet the prescribed reduced interconnected system accuracy requirement. 

The basic concept of abstracted reduction of $\llft(\env,\sys)$, as considered in \cite{Poort2024AbstractedReduction}, only considers one system connected to one environment. The current paper considers a far more generic problem setting of general interconnected systems $\llft(S,\sys_B)$ (see \autoref{ssec:sys_repr}), consisting of (many) interconnected subsystems. Whereas in the problem setting in \cite{Poort2024AbstractedReduction} only 
one abstraction error and one reduction error play a role, abstracted reduction of the more general interconnected systems $\llft(S,\sys_B)$ considered here introduces many more error sources, significantly challenging the analysis of the impact of abstraction errors and reduction errors on the overall accuracy of the reduced-order model. In addition, consideration of problems in which three or more subsystems $\ssys{j}$ play a role also allows for an alternative abstracted reduction approach, where the abstracted environment is composed by interconnecting abstracted subsystem models, instead of abstracting a high-order environment model. The current paper addresses these challenges, thereby enabling the abstracted reduction of general interconnected systems $\llft(S,\sys_B)$, for which the problem statement is formulated hereafter.

\subsection{Problem statement}
Our goal is to address the limitations of existing structure-preserving model reduction methods. Particularly, given the subsystems $\ssys{1}(s),\dots,\ssys{k}(s)$ and interconnection dynamics $S(s)$, constituting $\llft(S,\sys_B)$, we aim to reduce $\ssys{j}(s)$ to $\ssysr{j}(s)$ for all $j = 1,\dots, k$, such that
\begin{enumerate}
    \item stability is preserved, i.e., $\llft(S,\sysr_B)$ is well-posed and internally stable, in particular $\llft(S,\sysr_B) \in \RH$,
    \item the approximation is accurate in the sense that the approximation error $\ec$, as in \autoref{eq:ec_init_def}, is small.
\end{enumerate} 

Specifically, we consider the common problem setting of a relatively low-order interconnection structure $\S$ and multiple large subsystems $\ssys{j}$, such that $\sys_B$ is of significantly higher order than $\ssys{j}$ for all $j = 1,\dots, k$. The high order of $\sys_B$ makes that the application of existing structure-preserving reduction methods to $\llft(S,\sys_B)$ is infeasible (or comes at large computational cost).

%
%
\section{Abstracted reduction of interconnected systems}\label{sec:absred} 

\begin{figure}
    \subfloat[\small \label{fig:RGBred_envAbs}]{%
      \includegraphics[width=.48\linewidth]{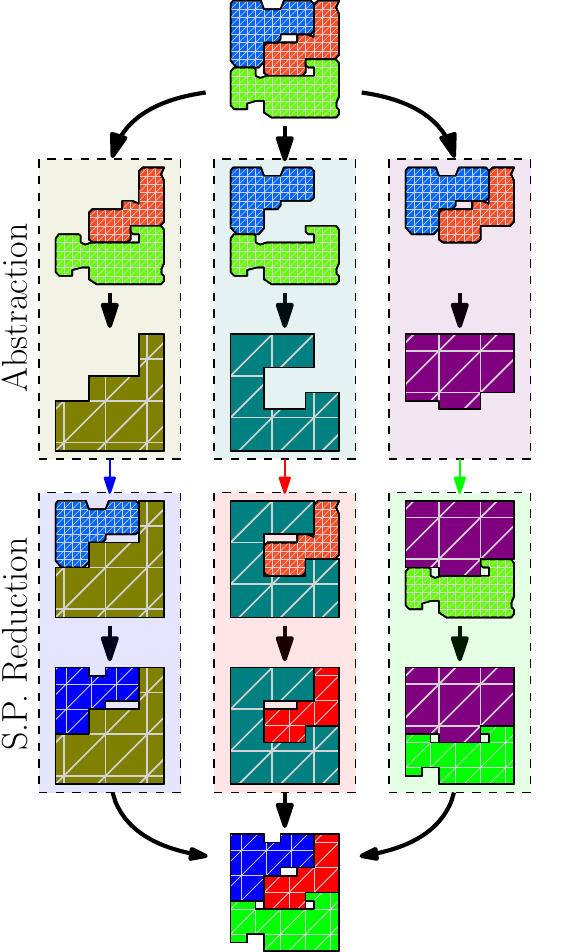}
    }
    \subfloat[\small\label{fig:RGBred_ssAbs}]{%
      \includegraphics[width=.48\linewidth]{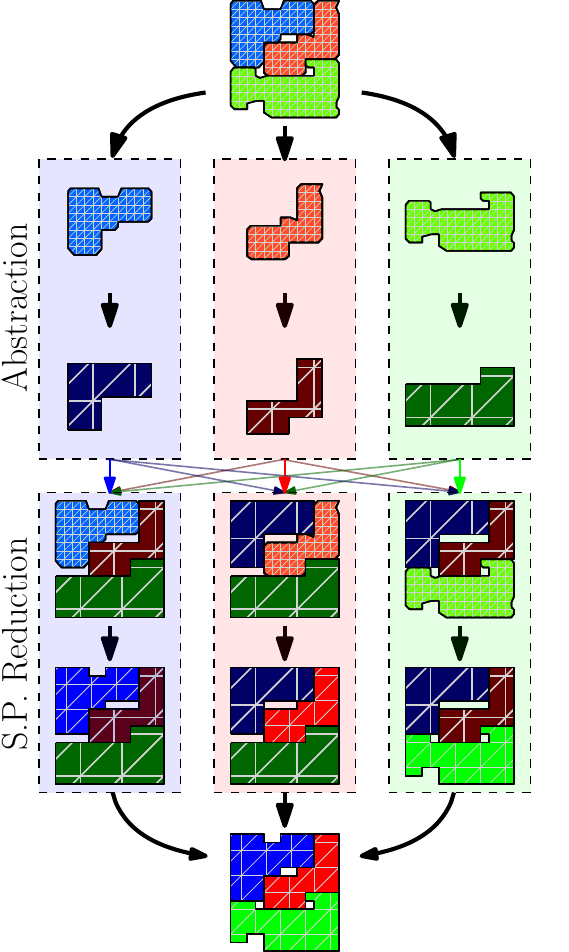}
    }\hspace{-2mm}
    \caption{Abstracted reduction using (a) environment abstraction (\autoref{alg:env_absred}), and (b) using subsystems abstraction (\autoref{alg:ss_absred}). } 
    \label{fig:RGBred_Abs_ill}
\end{figure}

The original framework of abstracted reduction, as described in \autoref{ssec:absred_prelim}, can be used to reduce every subsystem model $\ssys{j}$ within the interconnected system $\llft(\S,\sys_B)$ by adopting the environment-system representation of \autoref{fig:coupling_diag_env}. This approach is graphically represented by \autoref{fig:RGBred_envAbs}, where the structure-preserving reduction of a single subsystem within its abstracted environment can be identified as a single column. 

The precise steps to reduce all subsystem models $\ssys{j}$ using abstracted reduction, with separately abstracted environment models, are presented in the following algorithm.
\begin{alg} \label{alg:env_absred}
    \emph{Abstracted reduction 1 - environment abstraction}\\
    \textbf{Input:} Interconnection dynamics $S(s)$, transfer matrices $\ssys{j}(s)$ for $j=1,\dots,k$, abstraction and reduction orders $r_{\env,j}\leq n_{\env,j}$ and $r_{\sys,j}<n_{\sys,j}$, respectively, and weighting matrices $G^{j}_y\in\C^{p_j\times p_j}$ and $G_u^{j}\in\C^{m_j\times m_j}$.\\
    \textbf{Output:} Surrogate models $\ssysr{j}(s)$ for $j=1,\dots,k$ of reduced order $r_{\sys,j}$, such that $\llft(\S,\sysr_B)$ approximates $\llft(\S,\sys_B)$.
    
\begin{enumerate}[1),leftmargin = 5mm]
    \item \textbf{For all $j = 1, \dots, k$}
    \begin{enumerate}
        \item \textbf{Construction of $\senv{j}(s)$}. Construct the environments $\senv{j}(s)$, according to \autoref{def:senv}.
        \item \textbf{Abstraction of $\senv{j}(s)$}. Reduce $\senv{j}(s)$ to $\senva{j}(s)$ of order $r_{\env,j}$ in open loop.
        \item \textbf{Augmentation of $\senva{j}(s)$}. Augment the set of inputs and outputs of $\llft(\senva{j},\ssys{j})$ by incorporating $\ssys{j}$'s (weighted) inputs and outputs. This is equivalent to replacing $\senva{j}(s)$ with $\sfnva{j}(s)$, resulting in $\llft(\sfnva{j},\ssys{j})$, where
        \begin{equation} \label{eq:F_defs}
            \sfnva{j}(s) = \begin{bmatrix}    \sfnva{j}_{11} & \sfnva{j}_{12}\\ \sfnva{j}_{21} & \sfnva{j}_{22}   \end{bmatrix} 
            = \left[\begin{array}{cc|c} \senva{j}_{11} & O & \senva{j}_{12}\\ O &O &G_y^{j}\\ \hline \senva{j}_{21} & G^{j}_u & \senva{j}_{22}\end{array}\right].
        \end{equation}
        \item \textbf{Reduction of $\ssys{j}(s)$}. Use a structure-preserving reduction method to reduce $\llft(\sfnva{j},\ssys{j})$ to $\llft(\sfnva{j},\ssysr{j})$.
    \end{enumerate}
    \item \textbf{Connection of $\ssysr{j}(s)$}. Compose the reduced-order, interconnected system $\llft(\S,\sysr_B)$, with $\sysr_B \!= \!\diag(\ssysr{1},\shdots,\ssysr{k})$.
\end{enumerate}
\end{alg}

This framework of abstracted reduction reduces the computational cost of expensive structure-preserving reduction methods by first abstracting the environment using a computationally cheap, open-loop reduction method. In this way, the structure-preserving reduction of $\ssys{j}$ produces reduced subsystem models $\ssysr{j}$ that are relevant to the external input-output behaviour of $\llft(\senva{j},\ssys{j})$ instead of $\llft(\senv{j},\ssys{j})$. Thereby, abstracted reduction uses structure-preserving reduction to approximate the \emph{abstracted} coupled input-output behaviour where approximation of the actual behaviour is infeasible or too computationally expensive. 

\begin{rem}\label{rem:GyGu_influence}
    The augmentation of $\senva{j}$ to $\sfnva{j}$ with nonsingular weighting matrices $G^{j}_y$ and $G^{j}_u$ is beneficial for both the theoretical analysis of reduction error bounds in \autoref{sec:err_analysis} and for providing additional control over the accuracy distribution between $\llft(\senva{j},\ssysr{j})$ and $\ssysr{j}$. Reduction methods that approximate a model's input-output behavior naturally emphasize pairs with a high magnitude. Consequently, increasing the magnitudes of $G^{j}_y$ and $G^{j}_u$ enhances the accuracy of $\ssysr{j}$ but reduces the accuracy of $\llft(\senva{j},\ssysr{j})$ and vice versa.
\end{rem}

Besides the environment abstraction approach, as presented in \autoref{alg:env_absred}, it is also possible to generate abstracted subsystem models $\ssysa{j}$ and use these to construct $\senva{j}$. This approach of subsystem abstraction is visualized in \autoref{fig:RGBred_ssAbs} and introduced explicitly hereafter.

\begin{alg} \label{alg:ss_absred}
    \emph{Abstracted reduction 2 - subsystem abstraction}\\
    \textbf{Input:} Interconnection dynamics $S(s)$ and transfer matrices $\ssys{j}(s)$ for $j=1,\dots,k$, abstraction and reduction orders $r_{A,j}\leq n_{\sys,j}$ and $r_{\sys,j}<n_{\sys,j}$, respectively, and weighting matrices $G^{j}_y\in\C^{p_j\times p_j}$ and $G_u^{j}\in\C^{m_j\times m_j}$.\\
    \textbf{Output:} Surrogate models $\ssysr{j}(s)$ for $j=1,\dots,k$ of reduced order $r_{\sys,j}$, such that $\llft(\S,\sysr_B)$ approximates $\llft(\S,\sys_B)$.
    
    \begin{enumerate}[1),leftmargin = 5mm]
        \item \textbf{For all $j = 1, \dots, k$}
        \begin{enumerate}
            \item \textbf{Abstraction of $\ssys{j}(s)$}. Reduce $\ssys{j}(s)$ to $\ssysa{j}(s)$ of order $r_{A,j}$ in open loop.
        \end{enumerate}
        \item \textbf{For all $j = 1, \dots, k$}
        \begin{enumerate}
            \item \textbf{Construction of $\senva{j}(s)$}. Construct the abstracted environments $\senva{j}$ using $\S$ and $\ssysa{1},\dots,\ssysa{k}$, analogous to \autoref{def:senv}.
            \item \textbf{Augmentation of $\senva{j}(s)$}. Create the augmented system $\llft(\sfnva{j},\ssys{j})$, with $\sfnva{j}$ given as in \autoref{eq:F_defs}.
            \item \textbf{Reduction of $\ssys{j}(s)$}. Use a structure-preserving reduction method to reduce $\llft(\sfnva{j},\ssys{j})$ to $\llft(\sfnva{j},\ssysr{j})$.
        \end{enumerate}
        \item \textbf{Connection of $\ssysr{j}(s)$}. Compose the reduced-order, interconnected system $\llft(\S,\sysr_B)$, with $\sysr_B = \diag(\ssysr{1},\dots,\ssysr{k})$.
\end{enumerate}
\end{alg}

Given a certain abstraction order $r_{E,j}$, the environment abstraction approach of \autoref{alg:env_absred} typically results in higher accuracy of $\senva{j}$ than the subsystem abstraction approach of \autoref{alg:ss_absred} by not retaining the structure of $\senv{j}$. However, the subsystem abstraction approach of \autoref{alg:ss_absred} offers a higher level of modularity. Considering the design process of an interconnected system, where each subsystem is designed by a different design team, the approach of \autoref{alg:ss_absred} requires each team to provide a low-order approximate model of their subsystem $\ssysa{j}$, whereas \autoref{alg:env_absred} requires high-fidelity subsystem models $\ssys{j}$ to compose the full environments $\senv{j}$. In earlier stages of the design process, when most designs are not finalized, such approximate subsystem models $\ssysa{j}$ may already be available by exploiting previous revisions or preliminary knowledge of the subsystem. Therefore, \autoref{alg:ss_absred} may be most beneficial during (earlier stages of) the design process, whereas \autoref{alg:env_absred} presents the most reliable and efficient approach once all subsystem models are available.

Additionally, \autoref{alg:ss_absred} can use the acquired reduced, subsystem models $\ssysr{j}(s)$ as the abstracted model $\ssysa{j}(s)$ in the structure-preserving reduction of subsequent subsystem models. This use of the reduced subsystems as the abstracted subsystems is optional and only works if $r_{\sys,j}$ is an appropriate order of the abstraction, balancing accuracy and computational cost. If $r_{\sys,j}$ is indeed an appropriate order, it is even possible to iterate over step 2) in \autoref{alg:ss_absred}, setting $\ssysa{j}(s) = \ssysr{j}(s)$ after each iteration, which improves the accuracy of $\llft(\sfnva{j},\ssys{j})$ and thereby the accuracy of $\llft(\S,\sysr_B)$.

%
%
\section{Error analysis and the relation of error bounds}\label{sec:err_analysis}
To be able to guarantee the stability of $\llft(\S,\sysr_B)$, and to ensure a desired level of accuracy, we need to be able to relate the interconnected system reduction error $\ec$, as in \autoref{eq:ec_init_def}, to the introduced abstraction and reduction errors. To this end, we initially assume the abstraction and reduction errors to be known transfer functions and find explicit expressions for $\ec$, i.e., for both Algorithms \ref{alg:env_absred} and \ref{alg:ss_absred}, in \autoref{ssec:error_charac}. These expressions are subsequently used in \autoref{ssec:robperf} to also relate $\Hnrm{\infty}$-bounds on these errors using methods from robust performance analysis. These relations between error bounds form the basis for \autoref{sec:spec}, where a priori, sufficient conditions on the abstraction and reduction errors are formulated that guarantee the accuracy and stability of $\llft(\S,\sysr_B)$.

\subsection{Error relations} \label{ssec:error_charac}
We start by defining the error sources introduced in the environment and subsystem abstraction algorithms of \autoref{alg:env_absred} and \autoref{alg:ss_absred}. First, with environment abstraction (\autoref{alg:env_absred}), the environment models $\senv{j}$ are replaced by abstractions $\senva{j}$ for all $j = 1,\dots,k$, resulting in environment abstraction errors
\begin{equation}\label{eq:see_def}
    \see{j}(s) \coloneqq\senva{j}(s) - \senv{j}(s).
\end{equation}

In case of subsystem abstraction (\autoref{alg:ss_absred}) the abstracted environments are obtained by a composition of abstracted subsystem models $\ssysa{j}$ as $\senva{j} = \llft\big(\Sb{j},\ssysba{j}_B\big)$, analogous to \autoref{def:senv}. To this end, each subsystem model $\ssys{j}$ is reduced to $\ssysa{j}$, introducing the subsystem reduction error 
\begin{equation}\label{eq:sea_def}
    \sea{j}(s) \coloneqq\ssysa{j}(s) - \ssys{j}(s).
\end{equation}

We recall from Algorithms \ref{alg:env_absred} and \ref{alg:ss_absred} that the resulting abstract environment models $\senva{j}$ are subsequently augmented with matrices $G_u^{j}$ and $G_y^{j}$ to obtain $\sfnva{j}$. Augmentation does not affect the well-posedness and stability of the interconnection, as stated next.

\begin{lem}
    Consider the $p_j\times m_j$ transfer matrix $\ssys{j}$ and $(p_C+m_j)\times(m_C+p_j)$ transfer matrix $\senva{j}$, such that $\llft(\senva{j},\ssys{j})$ is well-posed and internally stable. Then, for any $\sfnva{j}$ as in \autoref{eq:F_defs}, with weighting matrices $G_u^{j}\in\C^{m_j\times m_j}$ and $G_y^{j}\in\C^{p_j\times p_j}$, $\llft(\sfnva{j},\ssys{j})$ is well-posed and internally stable.
\end{lem}
\begin{proof}
   The proof follows directly from \cite[Lemma~1]{Poort2024AbstractedReduction}.
\end{proof}

Next, the structure-preserving reduction of $\llft(\sfnva{j},\ssys{j})$ leads to the reduced subsystem models $\ssysr{j}$ and introduces the errors
\begin{equation}\label{eq:sef_def}
    \sef{j}(s) \coloneqq\llft\big(\sfnva{j}(s), \ssysr{j}(s)\big) - \llft\big(\sfnva{j}(s), \ssys{j}(s)\big).
\end{equation}
We emphasize that we work with the error $\sef{j}$ rather than $\ssysr{j}-\ssys{j}$ as $\sef{j}$ has relevance in the structure-preserving reduction of interconnected dynamics $\llft(\sfnva{j},\ssys{j})$. As such, it is expected to give a better indication of the quality of the reduction in the context of the interconnected system dynamics.

The abstraction and reduction steps, characterized through the errors in \autoref{eq:see_def} or \autoref{eq:sea_def} and \autoref{eq:sef_def}, respectively, ultimately lead to the reduction error on the interconnected system as
\begin{equation}\label{eq:ec_def}
    \ec(s) \coloneqq \llft\big(\S(s),\sysr_B(s)\big)- \llft\big(\S(s),\sys_B(s)\big).
\end{equation}

Our main goal in this section is to relate $\ec$ to the errors $\see{j}$, $\sea{j}$ and $\sef{j}$ for all $j = 1,\dots,k$. As $\ec(s)$ is fully determined by $\sysr_B(s)$, which is a block-diagonal connection of $\ssysr{j}(s)$, $j = 1,\dots,k$, we first state the following lemma on block-diagonal connections.
\begin{lem} \label{lem:diag_lft_order}
    Let $\ssys{1}(s)$, $\ssys{2}(s)$, $\senv{1}(s)$, $\senv{2}(s)$ be transfer function matrices, such that $\mathcal{F}(\senv{1},\ssys{1})$ and $\mathcal{F}(\senv{2},\ssys{2})$ are well-posed, where $\mathcal{F}$ represents either a lower- or upper-LFT. Then,
    \begin{multline}
            \mathcal{F}\left(\diag\big(\senv{1},\senv{2}\big),\ \diag\big(\ssys{1},\ssys{2}\big)\right) = \\ \diag\left(\mathcal{F}\big(\senv{1},\ssys{1}\big),\ \mathcal{F}\big(\senv{2},\ssys{2}\big)\right).
    \end{multline}
\end{lem}
\begin{proof}
    The proof follows from \autoref{def:llft_ulft_wellposed} and standard properties of the inversion and multiplication of block-diagonal matrices.
\end{proof}

\begin{figure}
    \centering
    \includegraphics{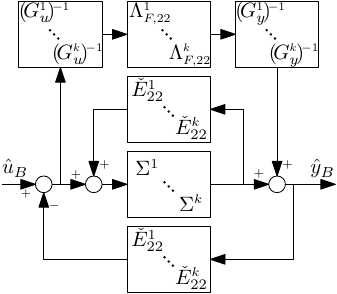}
    \caption{Schematic representation of $\sysr_B\coloneqq \diag(\ssysr{1},\dots,\ssysr{k})$, as in \autoref{eq:Eclem_sysrb}.}
    \label{fig:sysrB_expr}
\end{figure}

Using \autoref{lem:diag_lft_order}, we can express $\sysr_B\coloneqq \diag(\ssysr{1},\dots,\ssysr{k})$ in terms of the unreduced transfer functions, the abstracted environments $\senva{j}$ and the structure-preserving reduction errors $\sefs{j}{22}$, as also visualized in \autoref{fig:sysrB_expr}, in the following lemma.

\begin{lem}\label{lem:error_corr} 
    For $j = 1,\dots,k$, let $\ssys{j}(s)$, $\ssysr{j}(s)$ be transfer function matrices and let $\sfnva{j}(s)$ be as in \autoref{eq:F_defs} with square, invertible matrices $G_u^{j}$ and $G_y^{j}$, such that $\llft(\sfnva{j},\ssys{j})$ and $\llft(\sfnva{j},\ssysr{j})$ are well-posed with a difference $\sef{j}(s)$ as in \autoref{eq:sef_def}.
    Then,
    \begin{multline}\label{eq:Eclem_sysrb}
        \sysr_B = \ulft\left(\left[\begin{smallmatrix}- \enva_{B,22}&I\\I&O\end{smallmatrix}\right],\
            \sys_B(I-\enva_{B,22}\sys_B)^{-1} \right.\\  \left. + \left(G_y\right)^{-1} \efs{22} \left(G_u\right)^{-1} \right)
    \end{multline}
    is well-posed, where
    \begin{equation}\label{eq:enva_B22_def}
        \enva_{B,22}(s) = \diag(\enva_{22}^{1}(s),\,\dots,\, \enva_{22}^{k}(s))
    \end{equation}
    is the parallel connection of all $\enva_{22}^{j}$ and
    \begin{equation} \label{eq:efs22_def}
        \efs{22} = \llft\Big(\left[\begin{smallmatrix}  O&G_y\\G_u & \enva_{B,22}\!\end{smallmatrix}\right]\!, \sysr_B\Big) - \llft\Big(\left[\begin{smallmatrix}  O&G_y\\G_u & \enva_{B,22}\!\end{smallmatrix}\right]\!, \sys_B\Big),
    \end{equation}
    is the parallel connection of the $22$-partitions of all $\sef{j}$, related to the augmented inputs and outputs of $\llft(\sfnva{j},\ssys{j})$, where we use
    \begin{equation}
        G_u = \diag(G^{1}_u,\shdots,G^{k}_u), \quad G_y =    \diag(G^{1}_y,\shdots,G^{k}_y).        
    \end{equation}
\end{lem}
\begin{proof}
    The proof is based on \cite[Lemma~2]{Poort2024AbstractedReduction}, which provides a similar expression for a single subsystem. Expression \autoref{eq:Eclem_sysrb} can then be attained straightforwardly using \autoref{lem:diag_lft_order}. 
\end{proof}

Using the expression for $\sysr_B$ of \autoref{eq:Eclem_sysrb}, the interconnected system reduction error $\ec$ can be expressed as a function of $\sef{j}$ (i.e., the subsystem reduction errors) and $\senva{j}$ (i.e., the abstracted environment models). In case of environment abstraction (\autoref{alg:env_absred}), $\senva{j}$ depends on the error $\see{j}$, whereas for subsystem abstraction (\autoref{alg:ss_absred}) $\senva{j}$ depends on the errors $\sea{j}$. We will consider the expression for $\ec$ separately for environment and subsystem abstraction in Sections \ref{sssec:envabs} and \ref{sssec:sysabs}, respectively.\\

\subsubsection{Direct environment abstraction} \label{sssec:envabs} With environment abstraction, as treated in \autoref{alg:env_absred}, $\ec$ is determined by $\see{j}$ and $\sef{j}$ for all $j = 1,\dots,k$, as given in \autoref{eq:see_def} and \autoref{eq:seft_def}, respectively. Using \autoref{lem:error_corr}, we formulate the following expression for $\ec$.


\begin{figure}
    \centering
    \includegraphics{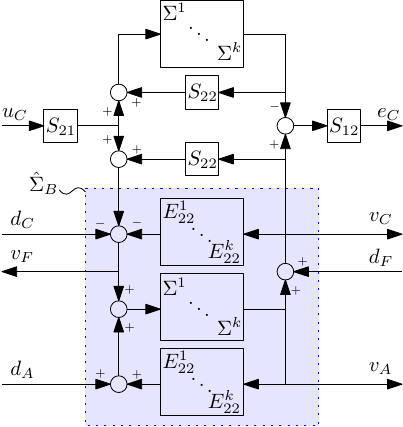}
    \caption{Schematic representation of $N_\env(s)$, as given in \autoref{eq:Nenv_def}, where $d_A$, $d_F$ and $d_C$ represent, respectively, the error sources from abstraction, reduction and connection (\autoref{alg:env_absred} steps 1.b, 1.d and 2) and $v_A$, $v_F$ and $v_C$ the related outputs.}
    \label{fig:env-Ec_LFT-Nexpr}
\end{figure}

\begin{thm}\label{th:ec_env_lft}
    Let $\sys_B(s)$, $\sysr_B(s)$ be transfer function matrices and let $\sfnva{j}(s)$ be as in \autoref{eq:F_defs}, satisfying \autoref{ass1}. Let $\see{j}(s)$ and $\sefs{j}{22}(s)$ be as in \autoref{eq:see_def} and \autoref{eq:sef_def}, respectively, and let 
    \begin{equation}\label{eq:seft_def}
         \seft{j}(s) \coloneqq \left(G_y^{j}\right)^{-1} \sefs{j}{22}(s) \left(G_u^{j}\right)^{-1}.
    \end{equation}
    Then, the error $\ec$ as in \autoref{eq:ec_def} satisfies
    \begin{multline}\label{eq:ec_env_lft}
            \ec = \ulft\Big(N_\env,\,\diag\big(\sees{1}{22},\,\dots,\,\sees{k}{22},\\
            \seft{1},\,\dots,\,\seft{k},\,\sees{1}{22},\,\dots,\,\sees{k}{22}\big)\Big),        
    \end{multline}
    where $N_\env(s)$ is visualized in \autoref{fig:env-Ec_LFT-Nexpr} and given by
    \begin{equation} \label{eq:Nenv_def}\centering
        N_\env =\left[\begin{array}{ccc|c}
        M & I+MZ & M & MS_{21}\hspace{-12mm}\\
        \hspace{-2mm}-ZM\!-\!I\hspace{-2mm}& \hspace{-2mm} Z(I\!+\!MZ)\hspace{-2mm} & \hspace{-2mm}ZM \hspace{-2mm}&\hspace{-1mm} (I\!+\!ZM)S_{21}\hspace{-14mm}\\
        -M & MZ & M & MS_{21}\hspace{-10mm}\\ \hline
        \hspace{-2mm}-S_{12}M \hspace{-2mm}& \hspace{-2mm}S_{12}(I\!+\!MZ) \hspace{-2mm}& \hspace{-1mm}S_{12}M \hspace{-3mm}& O\hspace{-12mm}
        \end{array}\right]
    \end{equation}
    and where
    \begin{align} \vspace{-5mm}
        Z(s) &= (S_{22}(s)- E_{B,22}(s)), \label{eq:Z_def}\\
        M(s) &= (I-\sys_B(s)S_{22}(s))^{-1}\sys_B(s). \label{eq:M_def}
    \end{align} 
\end{thm}
\begin{proof}
    The expression of $\ec$ as in \autoref{eq:ec_env_lft} can be verified intuitively by its graphical representation in \autoref{fig:env-Ec_LFT-Nexpr}. Starting from \autoref{fig:sysrB_expr}, which represents the expression for $\sysr_B$ of \autoref{eq:Eclem_sysrb}, we express each $\senva{j}_{22}$ as a parallel connection of $\senv{j}_{22}$ and $\sees{j}{22}$ and ``pull out'' the error terms $\sees{j}{22}$ and $\seft{j}$, resulting in the highlighted area of \autoref{fig:env-Ec_LFT-Nexpr}. This highlighted area, representing $\sysr_B$, is then interconnected with $\S$ to form $\llft(\S,\sysr_B)$. Following the definition of \autoref{eq:ec_def}, we then subtract $\llft(\S,\sys_B)$ to obtain $\ec$. The resulting diagram is given by \autoref{fig:env-Ec_LFT-Nexpr}, which has $\autoref{eq:Nenv_def}$ as its transfer function. This shows that, indeed, \autoref{eq:ec_env_lft} expresses $\ec$ as in \autoref{eq:ec_def}. The formal proof follows a similar line of reasoning as that of \cite[Theorem~1]{Poort2024AbstractedReduction}.
\end{proof}

With environment abstraction, $k$ abstraction errors $\sees{j}{22}$ and $k$ reduction errors $\seft{j}$ are introduced. Because each abstraction error $\sees{j}{22}$ appears twice, see \autoref{fig:env-Ec_LFT-Nexpr}, the expression of \autoref{eq:ec_env_lft} contains $3k$ error sources of which $2k$ are unique.\\

\subsubsection{Subsystem abstraction}\label{sssec:sysabs}
Alternatively, every $j$'th abstracted environment model $\senva{j}$ can be built from the interconnection dynamics $\S$ and the $k-1$ abstracted subsystem models $\ssysa{l}$, for all $l\in \{1,\dots,k\}\backslash\{j\}$, analogous to \autoref{def:senv} and \autoref{fig:env_expr}. This subsystem abstraction introduces the abstraction errors $\sea{j}(s)$ given by \autoref{eq:sea_def}, which we can relate to the interconnected system reduction error $\ec$ using the following theorem.

\begin{figure}
    \centering
    \includegraphics{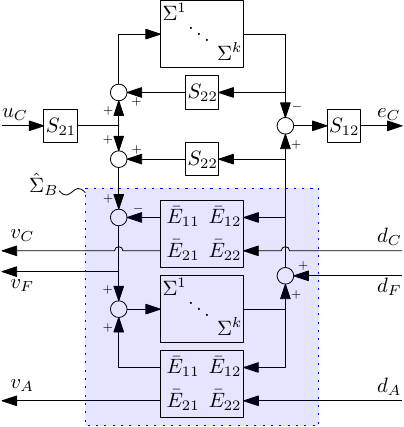}
    \caption{Schematic representation of $N_\sys(s)$, as given in \autoref{eq:Nss_def}, where $d_A$, $d_F$ and $d_C$ represent, respectively, the error sources from abstraction, reduction and connection (\autoref{alg:env_absred} steps 1.a, 2.c and 3) and $v_A$, $v_F$ and $v_C$ the related outputs.}
    \label{fig:ss-Ec_LFT-Nexpr}
\end{figure}

\begin{figure}
    \centering
    \includegraphics{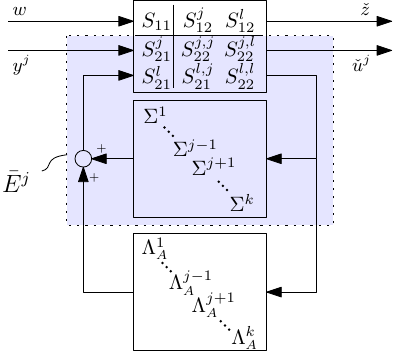}
    \caption{Schematic drawing of the $j$'th abstracted environment model $\senva{j}$ in case of subsystem abstraction, with the abstraction errors $\sea{j}$ pulled out. The extended environment model $\senvb{j}$, as given in \autoref{eq:ext_env}, is marked in blue.}
    \label{fig:extEnv_expr}
\end{figure}

\begin{thm}\label{th:ec_ss_lft}
    Let $\sys_B(s)$, $\sysr_B(s)$ be transfer function matrices and let $\fnva(s)$ be as in \autoref{eq:F_defs}, with $\enva(s)= \llft(\Sb{j},\ssysba{j}_B)$, satisfying \autoref{ass1}. Let $\sea{j}$ and $\seft{j}$ be as in \autoref{eq:sea_def} and \autoref{eq:seft_def}, respectively.
    Then, the error $\ec$ as in \autoref{eq:ec_def} satisfies
    \begin{multline}\label{eq:ec_ss_lft}
            \ec = \ulft\Big(N_\sys,\,\diag\big(\seab{1},\,\dots,\,\seab{k},\\
            \seft{1},\,\dots,\,\seft{k},\,\seab{1},\,\dots,\,\seab{k}\big)\Big), 
    \end{multline}
    where 
    \begin{equation}
        \seab{j} \coloneqq \diag\big(\sea{1},\dots,\sea{j-1},\sea{j+1},\dots,\sea{k}\big)
    \end{equation}
    and $N_\sys(s)$ is visualized in \autoref{fig:ss-Ec_LFT-Nexpr} and given by
    \begin{equation} \label{eq:Nss_def}\begingroup \setlength\arraycolsep{-0pt}
        N_\sys\! =\! \scalebox{0.82}{$\left[\begin{array}{ccc|c}
        \envb_{22}\!-\! \envb_{21}M\envb_{12} \hspace{2mm}&\hspace{2mm} \envb_{21}(I\!+\!MY) \hspace{2mm}&\hspace{2mm} \envb_{21}M\envb_{12} \qquad&\hspace{1mm} \envb_{21}MS_{21} \\
        -YM\envb_{12}\!-\!\envb_{12} \hspace{2mm}&\hspace{2mm} Y(I\!+\!MY) \hspace{2mm}&\hspace{2mm} YM\envb_{12} \hspace{5mm}&\hspace{1mm} (I\!+\!YM)S_{21} \\
        -\envb_{21}M\envb_{12} \hspace{2mm}&\hspace{2mm} \envb_{21}MY \hspace{2mm}&\hspace{2mm} \envb_{22}\!+\!\envb_{21}M\envb_{12} & \envb_{21}MS_{21} \\ \hline
        -S_{12}M\envb_{12} \hspace{2mm}&\hspace{2mm} S_{12}(I\!+\!MY) \hspace{2mm}&\hspace{2mm} S_{12}M\envb_{12} \hspace{5mm}&\hspace{1mm} O 
        \end{array}\right]$}\endgroup.
    \end{equation}
    For the expression of \autoref{eq:Nss_def}, we use $M(s)$ as in \autoref{eq:M_def}, $Y(s) = (S_{22}(s)-\envb_{11}(s))$ and $\envb(s)$, which collects all \emph{extended} environment models $\senvb{j}$, as
    \beqa\label{eq:parallel_extEnvs} 
        \envb &= \begin{bmatrix}
            \envb_{11} & \envb_{12}\\ \envb_{21} & \envb_{22} 
        \end{bmatrix}  \\
        &= \begin{bmatrix}
            \diag\!\left(\senvb{1}_{11},\dots, \senvb{k}_{11}\right) & \hspace{-2mm}   \diag\!\left(\senvb{1}_{12},\dots, \senvb{k}_{12}\right)\\
                \diag\!\left(\senvb{1}_{21},\dots, \senvb{k}_{21}\right) & \hspace{-2mm}  \diag\!\left(\senvb{1}_{22},\dots, \senvb{k}_{22}\right)
        \end{bmatrix},
    \eeqa
    where the extended environment models $\senvb{j}$ are given as
    \begin{equation}\label{eq:ext_env}
        \senvb{j} = 
        \begin{bmatrix} 
            S^{j,j}_{22} \!+\! S^{j,l}_{22}\ssysb{j}_BL S^{l,j}_{22} &   S^{j,l}_{22}(I\! +\! \ssysb{j}_BL S^{l,l}_{22})\\ 
            LS^{l,j}_{22} & L S^{l,l}_{22}
        \end{bmatrix},
    \end{equation}
    such that $\senva{j}_{22} = \llft(\senvb{j},\seab{j})$ as visualized in \autoref{fig:extEnv_expr}, and 
    \begin{equation}\label{eq:enva_b22_expr}
        \enva_{B,22} = \llft\big(\envb,\diag(\seab{1},\,\dots,\,\seab{k})\big).
    \end{equation}
    Note that in \autoref{eq:ext_env} $\ssysb{j}_B$ is as given in \autoref{eq:def_yb_ub_sysb}, $S^{a,b}_{22}$ with $a,b\in \{j,l\}$ as in \autoref{eq:def_Sb}, and $L(s)$ as
    \begin{equation}
        L(s) = \big(I-S^{l,l}_{22}(s)\ssysb{j}_B(s)\big)^{-1}.
    \end{equation} 
\end{thm}
\begin{proof}
    To be able to express $\sysr_B$ using \autoref{lem:error_corr}, we require an expression for $\enva_{B,22}$. However, with subsystem abstracted reduction, the abstraction error is added to each subsystem model as $\ssysa{j} = \ssys{j}+\sea{j}$, after which the abstracted environment models are composed as $\senva{j} = \llft(\Sb{j},\ssysba{j}_B)$. To extract the abstraction errors $\sea{j}$ from the expression for $\senva{j}_{22}$, the extended environments $\senvb{j}$ of \autoref{eq:ext_env} are defined such that $\senva{j}_{22} = \llft(\senvb{j},\seab{j})$. Using \autoref{lem:diag_lft_order}, we then collect all $\senva{j}_{22}$ in $\enva_{B,22}$ as in \autoref{eq:enva_b22_expr}.    
    We can then use \autoref{eq:enva_b22_expr} and \autoref{lem:error_corr} to express the interconnected system error as
    \begin{multline}\label{eq:ec_ss_expr}
        \ec = -\llft(\S,\sys_B) + \ulft\left(\left[\begin{smallmatrix}- \enva_{B,22}&I\\I&O\end{smallmatrix}\right],\right.\\
            \left. \sys_B(I-\enva_{B,22}\sys_B)^{-1}   + \diag(\seft{1},\dots,\seft{k}) \right).
    \end{multline}
    This can be rewritten to \autoref{eq:ec_ss_lft}, using the line of reasoning in the proof of \cite[Theorem~1]{Poort2024AbstractedReduction}.
\end{proof}

Note that, although the expression of \autoref{eq:ec_ss_lft} for the interconnected system reduction error in \autoref{th:ec_ss_lft} seems much more complex than \autoref{eq:ec_env_lft} in \autoref{th:ec_env_lft}, the theorems only differ in the expression for $\enva_{B,22}$. Whereas with environment abstraction $\enva_{B,22} = \env_{B,22} + \diag\big(\sees{1}{22},\,\dots,\,\sees{k}{22}\big)$, i.e., we have an additive error, subsystem abstraction requires the definition of extended environments $\envb$, such that $\enva_{B,22}$ can be expressed as in \autoref{eq:enva_b22_expr}.

With subsystem abstraction, each abstraction error $\sea{j}$ also influences all other $k-1$ abstracted environments $\senva{l}$, $l\in\{1,\dots,k\}\backslash\{j\}$. In the expression for $\ec$ of \autoref{eq:ec_ss_lft}, the environment errors also appear twice, such that each subsystem abstraction error $\sea{j}$ is repeated $2(k-1)$ times. Together with the $k$ reduction errors $\seft{j}$, this results in $2k^2-k$ error sources of which only $2k$ are unique.

\subsection{A robust performance perspective on error analysis}\label{ssec:robperf}
Using Theorems \ref{th:ec_env_lft} and \ref{th:ec_ss_lft}, the interconnected system reduction error $\ec$ can be expressed as a function of the introduced error sources (due to abstraction and subsystem reduction). However, we do not aim to relate the actual errors because they are generally unknown, but instead relate \emph{bounds} on these errors to guide the reduction approaches of Algorithms \ref{alg:env_absred} and \ref{alg:ss_absred} and to ensure stability and accuracy.

To this end, we assume a prescribed, weighted bound on the interconnected system reduction error $\ec(s)$ and aim to translate this bound to bounds on the reduction errors $\seft{j}(s)$ and bounds on the environment abstracted errors $\sees{j}{22}(s)$ (in case of \autoref{alg:env_absred}) or the subsystem abstracted errors $\sea{j}(s)$ (in case of \autoref{alg:ss_absred}). Specifically, we prescribe weighting functions $V_C(s)\in \RH$ and $W_C(s) \in \RH$ and restrict $\ec \in \uec$, where 
\begin{equation} \label{eq:uec_def}
    \uec  \coloneqq\big\{\ec \in \RH\ \big| \ \|V_C \ec W_C\|_\infty < 1 \big\}.
\end{equation}
Given this specification on the accuracy of the overall interconnected system dynamics, we aim to find similar specifications for abstraction and reduction errors $\seft{j}$ and $\sees{j}{22}$ or $\sea{j}$, by restricting them, respectively, to 
\begin{equation}\label{eq:acc_spec_def}
    \begin{aligned}
        \sueft{j} & \coloneqq\big\{\seft{j}\ \big| \ \|(\sW{j}_F)^{-1} \seft{j} (\sV{j}_F)^{-1}\|_\infty \leq 1 \big\},\\
        \suees{j}{22} & \coloneqq\big\{\sees{j}{22}\ \big|  \ \|(\sW{j}_E)^{-1} \sees{j}{22} (\sV{j}_E)^{-1}\|_\infty \leq 1 \big\},  \\  
        \suea{j} & \coloneqq\big\{\sea{j}\ \big|\  \|(\sW{j}_A)^{-1} \sea{j} (\sV{j}_A)^{-1}\|_\infty \leq 1 \big\},
    \end{aligned}
\end{equation}
which are defined by the bistable weighting functions $\sV{j}_\env$, $\sW{j}_\env,\,\sV{j}_A,\,\sW{j}_A,\,\sV{j}_F,\,\sW{j}_F,\,\sV{j}_C,\,\sW{j}_C\in \RH$ of appropriate dimensions. 

Our goal is now as follows: determine sets $\sueft{j}$ and $\suees{j}{22}$ or $\suea{j}$ (by appropriate choices of weighting functions), such that $\ec \in \uec$ is guaranteed by $\seft{j}\in \sueft{j}$ and $\sees{j}{22}\in \suees{j}{22}$ or $\sea{j}\in \suea{j}$. This relation between \emph{high-level} accuracy requirements (on the interconnected system reduction error) and \emph{low-level} requirements (on the underlying error sources) originates from techniques from robust performance analysis (see e.g. \cite{Zhou1998EssentialsControl}). It was first applied in the field of model reduction in \cite{Janssen2024ModularPerspective} for the open-loop reduction of the separate subsystem models, as illustrated in \autoref{fig:indepSSRed}. A similar approach will be used here to relate accuracy requirements within the presented framework of abstracted reduction.

To enable the use of these techniques for the problem setting in this paper, we first combine the error specifications of \autoref{eq:uec_def} and \autoref{eq:acc_spec_def} by defining the block-diagonal transfer function matrices
\beqa\label{eq:VW_env_def}
    \scalebox{0.86}{$\bm{W}_\env\,$}
    & \scalebox{0.86}{$=\diag\big(\sW{1}_\env,\shdots,\sW{k}_\env,\sW{1}_F,\shdots,\sW{k}_F,\sW{1}_\env,\shdots,\sW{k}_\env,W_C\big)$}, \\
    \scalebox{0.86}{$\bm{V}_\env\,$}
    &
    \scalebox{0.86}{$= \diag\big(\sV{1}_\env,\shdots,\sV{k}_\env,\sV{1}_F,\shdots,\sV{k}_F,\sV{1}_\env,\shdots,\sV{k}_\env,V_C\big)$}
\eeqa
for the case of environment abstraction. In case of subsystem abstraction, we instead use the transfer function matrices
\beqa\label{eq:VW_sys_def}
    \scalebox{0.86}{$\bm{W}_\sys\,$}
    & \scalebox{0.86}{$=\diag\big(\sWb{1}_A,\shdots,\sWb{k}_A,\sW{1}_F,\shdots,\sW{k}_F,\sWb{1}_A,\shdots,\sWb{k}_A,W_C\big)$},\\    
    \scalebox{0.86}{$\bm{V}_\sys\,$}
     & 
     \scalebox{0.86}{$=\diag\big(\sVb{1}_A,\shdots,\sVb{k}_A,\sV{1}_F,\shdots,\sV{k}_F,\sVb{1}_A,\shdots,\sVb{k}_A,V_C\big)$},
\eeqa
with
\beqa
    \sWb{j}_A &= \diag\big(\sW{1}_A,\dots,\sW{j-1}_A,\sW{j+1}_A,\dots,\sW{k}_A\big),\\
    \sVb{j}_A &= \diag\big(\sV{1}_A,\dots,\sV{j-1}_A,\sV{j+1}_A,\dots,\sV{k}_A\big).
\eeqa

For the case of environment abstraction, we also define the nonunique, nonsingular matrices $T_{\env,\ell}$ and $T_{\env,r}$ that sort all error terms of \autoref{eq:ec_env_lft} as
\begin{gather}
    \bm{\Lambda}_\env = T_{\env,\ell} \,\bm{\Lambda}_{\env,s}\, T_{\env,r},\quad \text{where}\\ \nonumber
    \scalebox{0.86}{$
    \begin{gathered}
    \bm{\Lambda}_\env = \diag\big(\sees{1}{22},\shdots,\sees{k}{22},\seft{1},\shdots,\seft{k},\sees{1}{22},\shdots,\sees{k}{22},\ec\big),\\
    \bm{\Lambda}_{\env,s} = \diag\big(\sees{1}{22},\sees{1}{22},\shdots,\sees{k}{22},\sees{k}{22},\seft{1},\shdots,\seft{k},\ec\big).
    \end{gathered}
    $}
\end{gather}

Analogously, for subsystem abstraction, we define the matrices $T_{\sys,\ell}$ and $T_{\sys,r}$ that sort all error terms of \autoref{eq:ec_ss_lft} as
\begin{gather}
    \bm{\Lambda}_\sys = T_{\sys,\ell} \,\bm{\Lambda}_{\sys,s}\, T_{\sys,r},\quad \text{where}\\
    \scalebox{0.86}{$
    \begin{gathered} \nonumber
        \bm{\Lambda}_\sys = \diag\big(\seab{1},\shdots,\seab{k},\seft{1},\shdots,\seft{k},\seab{1},\shdots,\seab{k},\ec\big)\\ \nonumber
        \bm{\Lambda}_{\sys,s} = \diag\big(\sea{1},\shdots,\sea{1},\shdots,\sea{k},\shdots,\sea{k},\seft{1},\shdots,\seft{k},\ec\big).        
    \end{gathered}
    $}
\end{gather}

These sorting matrices are used to (more easily) define the sets of scaling matrices $\mathbb{D}_\env$ and $\mathbb{D}_\sys$ in \autoref{eq:D_env_def} and \autoref{eq:D_sys_def}, respectively. We can now formulate the following sufficient condition for $\ec \in \uec$.

\begin{figure*}\centering 
    \begin{cequation} \label{eq:D_env_def}
            \mathbb{D}_{\env} = \left\{ (D_\ell, D_r) \ \left| \ \begin{gathered} D_\ell = T_{\env,\ell} S_\ell T_{\env,\ell}^\top, \ D_r =  T_{\env,r} S_r T_{\env,r}^\top\\
                S_\ell = \diag\big(S^{1}\!\otimes\! I_{p_1},\, \shdots,\,S^{k}\!\otimes\! I_{p_k}, d^{1}I_{m_1},\,  \shdots,\,  d^{k}I_{m_k},I_{m_C}\big), \\
                S_r = \diag\big(S^{1}\!\otimes\! I_{m_1},\,  \shdots,\,  S^{k}\!\otimes\! I_{m_k},\,  d^{1}I_{p_1},\shdots,\,  d^{k}I_{p_k},I_{p_C}\big), \\
                S^{j} \in \C^{2\times 2},\
                S^{j} = (S^{j})^H \succ 0,\
                d^{j} \in \R_{>0},\  \forall \ j = 1,\dots, k.
            \end{gathered}
            \right.\right\} 
    \end{cequation}
    \begin{cequation}\label{eq:D_sys_def}
            \mathbb{D}_{\sys} = \left\{ (D_\ell, D_r) \ \left| \ \begin{gathered} D_\ell = T_{\sys,\ell} S_\ell T_{\sys,\ell}^\top, \ D_r =  T_{\sys,r} S_\ell T_{\sys,r}^\top\\
                S_\ell = \diag\big(S^{1}\!\otimes\! I_{m_1},\,  \shdots,\,  S^{k}\!\otimes\! I_{m_k},\,  d^{1}I_{m_1},\, \shdots,\,   d^{k}I_{m_k},I_{m_C}\big), \\
                S_r = \diag\big(S^{1}\!\otimes\! I_{p_1},\,  \shdots,\,S^{k}\!\otimes\! I_{p_k},\,  d^{1}I_{p_1},\,  \shdots,\,  d^{k}I_{p_k},I_{p_C}\big), \\
                S^{j} \in \C^{2(k-1)\times 2(k-1)},\
                S^{j} = (S^{j})^H \succ 0,\
                d^{j} \in \R_{>0},\ \forall \ j = 1,\dots, k.
            \end{gathered}
            \right.\right\} 
    \end{cequation}
\end{figure*}

\begin{thm} \label{thm:req_validation}
    Consider the transfer functions $\sys_1,\dots,\sys_k$, $S$ and $\llft(S,\sys_B)$ satisfying \autoref{ass1}, which is reduced to $\llft(S,\sysr_B)$ using either
    \begin{itemize}
        \item {environment abstraction:} 
        let $N = N_\env$, $\bm{W} = \bm{W}_\env$, $\bm{V} = \bm{V}_\env$ and $\mathbb{D} = \mathbb{D}_\env$ as given in \autoref{eq:Nenv_def}, \autoref{eq:VW_env_def} and \autoref{eq:D_env_def}, respectively, and let $\sees{j}{22}\in \RH$ and $\seft{j}\in \RH$ for all $j=1,\dots,k$, or
        \item {subsystem abstraction:}
        let $N= N_\sys$, $\bm{W} = \bm{W}_\sys$, $\bm{V}= \bm{V}_\sys$ and $\mathbb{D} = \mathbb{D}_\sys$ as given in \autoref{eq:Nss_def}, \autoref{eq:VW_sys_def} and \autoref{eq:D_sys_def}, respectively and let $\sea{j}\in \RH$ and $\seft{j}\in \RH$ for all $j=1,\dots,k$.
    \end{itemize}
    Assume that $\bm{W}(s)$ and $\bm{V}(s)$ are bistable and biproper and let the requirements $\suees{j}{22},\ \suea{j},\ \sueft{j}$ and $\uec$ be given as in \autoref{eq:acc_spec_def} and \autoref{eq:uec_def}, respectively. If there exist $(D_\ell,D_r)\in \mathbb{D}$, such that
    \begin{equation}\label{eq:req_LMI}\hspace{-1mm}
        \mathcal{N}(i\omega) D_r\mathcal{N}^H(i\omega)\preceq D_\ell \quad \forall \,\omega \in \R, 
    \end{equation}
   with $\mathcal{N} = \bm{V}N\bm{W}$, then it follows that with
   \begin{itemize}
       \item environment abstraction, if $\seft{j}\in \sueft{j}$ and $\sees{j}{22}\in \suees{j}{22}$,
       \item subsystem abstraction, if $\seft{j}\in \sueft{j}$ and $\sea{j}\in \suea{j}$,
   \end{itemize}
   $\llft(S,\sysr_B)$ is well-posed and internally stable and the coupled error dynamics satisfies
   \begin{equation}\label{eq:ec_req}
        \ec\in \uec.
    \end{equation}
\end{thm}
\begin{proof}
The proof follows a similar line of reasoning as that in \cite[Theorem~2]{Poort2024AbstractedReduction} and is omitted for the sake of brevity.
\end{proof}

\autoref{thm:req_validation} connects high-level and low-level accuracy requirements, which are all defined by $\bm{V}$ and $\bm{W}$, through the inequality of \autoref{eq:req_LMI}. Explicitly, \autoref{thm:req_validation} can be used to check, a priori, whether satisfaction of certain low-level requirements $\sueft{j}$ and $\suees{j}{22}$ or $\suea{j}$ guarantees $\ec$ to satisfy the high-level requirement $\uec$. Note that the satisfaction of such accuracy requirements also guarantees the stability of the error dynamics (and thus of the reduced model) by assuming that $\bm{V}$ and $\bm{W}$ are bistable. As $\bm{V}$ and $\bm{W}$ are chosen by the user, they can always be chosen to satisfy this bistability assumption.

%
%
\section{Optimizing accuracy specifications for robust abstracted reduction}\label{sec:spec}
In the following, we exploit \autoref{thm:req_validation} to not only check whether certain low-level requirements $\sueft{j}$ and $\suees{j}{22}$ or $\suea{j}$ guarantee $\ec\in\uec$, but to find the \emph{least strict} low-level requirements that guarantee $\ec\in\uec$, such that we can reduce as far as possible. To this end, we prescribe some high-level accuracy specification by the set $\uec$ and formulate an optimization problem to maximize the set size of the abstraction accuracy specification ($\suees{j}{22}$ or $\suea{j}$) and reduction accuracy specification ($\sueft{j}$). These low-level specifications can then be used to reduce all subsystems and abstract all environments in a systematic manner and as far as possible, such that the reduced interconnected model still admits the prescribed error bound $\uec$ and is stable.

\subsection{Optimized low-level accuracy specifications}
 To efficiently determine appropriate low-level accuracy specifications, as in \autoref{eq:acc_spec_def}, we prescribe certain weighting matrices and aim to find scalar $\Hnrm{\infty}$-bounds on the weighted error systems that guarantee a certain weighted $\Hnrm{\infty}$-bound on the interconnected error system $\ec(s)$. Specifically, we prescribe the weighting matrices $\bm{W}_E(s)$ and $\bm{\Vc}_E(s)$ or $\bm{W}_\sys(s)$ and $\bm{\Vc}_\sys(s)$, where we define $\bm{\Vc}_E(s)$ and $\bm{\Vc}_\sys(s)$ as 
\begin{equation}\label{eq:VcE_def}\mbox{\small $
   \begin{aligned}
       \bm{\Vc}_\env(s) &\coloneqq \diag\big(\sVc{1}_\env,\shdots,\sVc{k}_\env,\sVc{1}_F,\shdots,\sVc{k}_F,\sVc{1}_\env,\shdots,\sVc{k}_\env,V_C\big),\\
       &=\bm{\eb}_\env^{-1} \bm{V}_\env(s)
    \end{aligned}$}  
\end{equation} 
and
\begin{equation}\label{eq:VcS_def}\mbox{\small $
   \begin{aligned}
       \bm{\Vc}_\sys(s) &\coloneqq \diag\big(\sVbc{1}_A,\shdots,\sVbc{k}_A,\sVc{1}_F,\shdots,\sVc{k}_F,\sVbc{1}_A,\shdots,\sVbc{k}_A,V_C\big),\\
       &=\bm{\eb}_\sys^{-1} \bm{V}_\sys(s),
    \end{aligned}    $}     
\end{equation} 
where 
\begin{align} \label{eq:ebb_env_def}
    \bm{\eb}_\env &= \diag\big(\sebe{1},\shdots,\sebe{k},\sebf{1},\shdots,\sebf{k},\sebe{1},\shdots,\sebe{k},1\big),\\
    \bm{\eb}_\sys &= \diag\big(\sebba{1},\shdots,\sebba{k},\sebf{1},\shdots,\sebf{k},\sebba{1},\shdots,\sebba{k},1\big),\label{eq:ebb_sys_def}
\end{align}
with
\begin{gather}
    \sebba{j} = \diag\big(\seba{1},\dots,\seba{j-1},\seba{j+1},\dots,\seba{k}\big),\\
    \sebe{j},\, \seba{j},\, \sebf{j} \in \R_{>0}, \text{ for } j = 1,\dots, k.
\end{gather}

The weighting matrices $\bm{W}_E(s)$ and $\bm{\Vc}_E(s)$ or $\bm{W}_\sys(s)$ and $\bm{\Vc}_\sys(s)$ effectively prescribe weighted $\Hnrm{\infty}$-bounds given by
\begin{equation} \label{eq:eb_def}
    \begin{aligned}
        &\|\big(\sW{j}_\env\big)^{-1} \sees{j}{22} \big(\sVc{j}_\env\big)^{-1}\big\|_\infty \leq \sebe{j}, \\
        &\|\big(\sWb{j}_A\big)^{-1} \sea{j} \big(\sVbc{j}_A\big)^{-1}\big\|_\infty \leq \seba{j}, \\
        &\|\big(\sW{j}_F\big)^{-1} \seft{j} \big(\sVc{j}_F\big)^{-1}\big\|_\infty \leq \sebf{j},
    \end{aligned}
\end{equation}
such that we can maximize $\sebe{j}$, $\seba{j},$ and $\sebf{j}$ to find the most lenient error bounds on $\sees{j}{22}$, $\sea{j},$ and $\seft{j}$. This maximization is performed using the following optimization problem.

\begin{thm}\label{th:opt_alg_TD_infty}
    Consider the setting as introduced in \autoref{thm:req_validation}. Instead of $\bm{V}$, let $\bm{\Vc}$ be either $\bm{\Vc}_\env$ or $\bm{\Vc}_\sys$, as defined in \autoref{eq:VW_env_def} and \autoref{eq:VW_sys_def}, for environment or subsystem abstraction, respectively. Then, $\bm{V}$ is defined by the variable $\bm{\eb}$, adopting the structure of either $\bm{\eb}_E$ or $\bm{\eb}_\sys$, as specified in \autoref{eq:ebb_env_def} and \autoref{eq:ebb_sys_def}.
Consider now the optimization problem where we use $P = \bm{\Vc} N \bm{W}$.
    \begin{align}\label{eq:opt_alg_TD_infty}
        \text{\emph{given} }\ & \bm{\Vc},\,\bm{W}\\
        \text{\emph{maximize} }\ & \|\bm{\eb}\| \nonumber\\
        \text{\emph{subject to} }\ & 
        \bm{\eb}^2 \, P(i\omega) D_rP^H(i\omega)\preceq D_\ell \ \forall \,\omega\in\R,  \nonumber\\
        &\ (D_\ell,D_r)\in \mathbb{D}. \nonumber
    \end{align}
    If $\bm{\eb}$ (defining $\bm{V}$ and thereby the accuracy specifications) is a feasible solution to \autoref{eq:opt_alg_TD_infty}, then it follows that with
   \begin{itemize}
       \item environment abstraction, if $\seft{j}\in \sueft{j}$ and $\sees{j}{22}\in \suees{j}{22}$,
       \item subsystem abstraction, if $\seft{j}\in \sueft{j}$ and $\sea{j}\in \suea{j}$,
   \end{itemize}
   $\llft(S,\sysr_B)$ is well-posed and internally stable and the coupled error dynamics satisfies
   \begin{equation}\label{eq:ec_req2}
        \ec\in \uec.
    \end{equation}
\end{thm}
\begin{proof}
    The proof follows from \autoref{thm:req_validation}, where $\bm{V}$ is replaced by $\bm{\eb\Vc}$. Substitution of $\bm{V}=\bm{\eb}\bm{\Vc}$ into the inequality condition of \autoref{eq:opt_alg_TD_infty} gives the condition of \autoref{eq:req_LMI}.
\end{proof}

\begin{rem} \label{rem:iterative_eb-Dr}
    The matrix inequality of \autoref{eq:opt_alg_TD_infty} is nonlinear in the unknown $\bm{\eb}$ and $D_r$, but can be solved efficiently by iteratively solving for $\bm{\eb}$ and $D_r$, as shown in \cite{Janssen2023ModularApproach}. Instead of evaluating all $\omega\in \R$, one can efficiently determine the supremum using the method of \cite{Bruinsma1990AMatrix}.
\end{rem}

\begin{rem} \label{rem:GyGu_influence2}
    The resulting specification $\sueft{j}$ determines how large the reduction error $\sefs{j}{22} = G_y^{j} \seft{j} G_u^{j}$ is allowed to be, which thus depends on the choice of $G_u^{j}$ and $G_y^{j}$. In \autoref{rem:GyGu_influence}, it was already mentioned that the magnitudes of these weighting matrices pose a trade-off between the required reduction order and the accuracy of $\llft(\S,\sysr_B)$. Specifically, larger magnitudes of $G_u^{j}$ and $G_y^{j}$ emphasize the accuracy of $\ssysr{j}$, thereby lowering $\seft{j}$ for a given $r_{\sys,j}$, whereas lower magnitudes emphasize the accuracy of $\llft(\senva{j},\ssysr{j})$, thereby reducing $\ec$. Low magnitudes of $G_u^{j}$ and $G_y^{j}$ therefore tend to result in more conservatism of $\ec$ with respect to $\uec$.
\end{rem}

\subsection{Robust abstracted reduction}
\autoref{th:opt_alg_TD_infty} helps to find the most lenient accuracy specifications $\suees{j}{22}$, $\suea{j}$ and $\sueft{j}$ that guarantee the stability of $\llft(\S,\sysr_B)$ and $\ec\in\uec$. Any environment or subsystem abstraction and subsystem reduction that meets these error specifications will ensure $\llft(\S,\sysr_B)$ to be stable and $\ec\in\uec$. Thus, by first determining such specifications and then abstracting and reducing to meet all accuracy specifications, we achieve a systematic approach to the abstracted reduction frameworks of Algorithms \ref{alg:env_absred} and \ref{alg:ss_absred}. We will first present this systematic framework, called \emph{robust abstracted reduction}, for the case of environment abstraction.
\vspace{3mm}

\begin{alg} \label{alg:robabsred_env}
    \emph{Robust Abstracted Reduction 1 - environment abstraction}\\
    \textbf{Input:} Transfer matrices $S(s)$ and $\ssys{j}(s)$ for $j = 1,\dots, k$, constituting $\senv{j}$ and $\llft(\S,\sys_B)$, satisfying \autoref{ass1}, bistable and biproper $\bm{\Vc}_\env$ and $\bm{W}_\env$ as in \autoref{eq:VcE_def} and \autoref{eq:VW_env_def} and nonsingular weighting matrices $G^{j}_y\in\C^{p_j\times p_j}$ and $G_u^{j}\in\C^{m_j\times m_j}$.\\
    \textbf{Output:} Surrogate models $\ssysr{j}(s)$ of reduced order $r_{\sys,j}$, such that $\llft(\S,\sysr_B)$ is well-posed, internally stable and $\ec \in \uec$ as in \autoref{eq:uec_def}.
\begin{enumerate}
    \item \textbf{Optimization} Solve the optimization problem given in \autoref{th:opt_alg_TD_infty} to attain specifications $\sueft{j}$ and $\suees{j}{22}$.
    \item \textbf{For all $j = 1,\dots,k$}
    \begin{enumerate}
        \item \textbf{Abstraction of $\senv{j}$}. Reduce $\senv{j}$ to $\senva{j}$ of the lowest order $r_{\env,j}$, such that $\sees{j}{22}\in \suees{j}{22}$.
        \item \textbf{Augmentation of $\senva{j}$}. Augment the inputs and outputs of $\llft(\senva{j},\ssys{j})$, resulting in $\llft(\sfnva{j},\ssys{j})$, with $\sfnva{j}$ as in \autoref{eq:F_defs}.
        \item \textbf{Reduction of $\ssys{j}(s)$}. Reduce $\ssys{j}$ to $\ssysr{j}$ of the lowest order $r_{\sys,j}$, such that $\seft{j} \in \sueft{j}$, where \\
        {\small \begin{equation}  \hspace{-8mm}
            \seft{j} = \llft\left(\left[\begin{smallmatrix}O &I\\ I & \senva{j}_{22} \end{smallmatrix}\right], \ssysr{j}\right) - \llft\left(\left[\begin{smallmatrix}O &I\\ I & \senva{j}_{22} \end{smallmatrix}\right], \ssys{j}\right).
        \end{equation}}
    \end{enumerate}
    \item \textbf{Connection of $\ssys{j}$}. Compose the reduced, interconnected system model $\llft(\S,\sysr_B)$.
\end{enumerate}
\end{alg}

A similar systematic framework can be formulated based on \autoref{th:opt_alg_TD_infty} and \autoref{alg:ss_absred} as follows.
\begin{alg} \label{alg:robabsred_ssys}
    \emph{Robust Abstracted Reduction 1 - subsystem abstraction}\\
    \textbf{Input:} Transfer matrices $S(s)$ and $\ssys{j}(s)$ for $j = 1,\dots, k$, constituting $\senv{j}$ and $\llft(\S,\sys_B)$, satisfying \autoref{ass1}, bistable and biproper $\bm{\Vc}_\sys$ and $\bm{W}_\sys$ as in \autoref{eq:VcS_def} and \autoref{eq:VW_sys_def} and nonsingular weighting matrices $G^{j}_y\in\C^{p_j\times p_j}$ and $G_u^{j}\in\C^{m_j\times m_j}$.\\
    \textbf{Output:} Surrogate models $\ssysr{j}(s)$ of reduced order $r_{\sys,j}$, such that $\llft(\S,\sysr_B)$ is well-posed, internally stable and $\ec \in \uec$ as in \autoref{eq:uec_def}.
\begin{enumerate}
    \item \textbf{Optimization} Solve the optimization problem given in \autoref{th:opt_alg_TD_infty} to attain specifications $\sueft{j}$ and $\suea{j}$.
    \item \textbf{For all $j = 1,\dots,k$}
    \begin{enumerate}
        \item \textbf{Abstraction of $\ssys{j}$}. Reduce $\ssys{j}$ to $\ssysa{j}$ of the lowest order $r_{A,j}$, such that $\sea{j}\in \suea{j}$.
    \end{enumerate}
    \item \textbf{For all $j = 1,\dots,k$}
    \begin{enumerate}
        \item \textbf{Construction of $\senva{j}$}. Construct the abstracted environment models $\senva{j}$ using $\S$ and all $k$ abstracted subsystem models $\ssysa{j}$, analogous to \autoref{def:senv}.
        \item \textbf{Augmentation of $\senva{j}$}. Augment the inputs and outputs of $\llft(\senva{j},\ssys{j})$, resulting in $\llft(\sfnva{j},\ssys{j})$, with $\sfnva{j}$ as in \autoref{eq:F_defs}.
        \item \textbf{Reduction of $\ssys{j}(s)$}. Reduce $\ssys{j}(s)$ to $\ssysr{j}(s)$ of the lowest order $r_{\sys,j}$, such that $\seft{j} \in \sueft{j}$, where 
        {\small \begin{equation} \hspace{-8mm}
            \seft{j} = \llft\left(\left[\begin{smallmatrix}O &I\\ I & \senva{j}_{22} \end{smallmatrix}\right], \ssysr{j}\right) - \llft\left(\left[\begin{smallmatrix}O &I\\ I & \senva{j}_{22} \end{smallmatrix}\right], \ssys{j}\right).
        \end{equation}}
    \end{enumerate}
    \item \textbf{Connection of $\ssys{j}$}. Compose the reduced, interconnected system model $\llft(\S,\sysr_B)$.
\end{enumerate}
\end{alg}

\begin{rem}
    Algorithms \ref{alg:robabsred_env} and \ref{alg:robabsred_ssys} determine the error specifications $\suees{j}{22}$ or $\suea{j}$ and $\sueft{j}$ only once at the start of the algorithm. However, it is always possible to update error specifications once additional knowledge of the imposed error sources is attained. For example, with \autoref{alg:robabsred_env}, when all environments $\senv{j}$ are replaced by $\senva{j}$ and $\sees{j}{22}$ is known exactly, it is possible to redetermine $\sueft{j}$, which will be a larger set than before due to the conservatism of $\sees{j}{22}$ with respect to $\suees{j}{22}$. Such a specification update thus allows for further reduction of the remaining, unreduced models, but comes at the additional cost of rerunning the optimization of \autoref{th:opt_alg_TD_infty}. 
\end{rem}

Besides the accuracy and stability guarantee of Algorithms \ref{alg:robabsred_env} and \ref{alg:robabsred_ssys}, one of the most important advantages is their systematic nature. Whereas meeting a high-level error specification typically requires some trial and error with low-level or structure-preserving reduction approaches, these abstracted reduction frameworks clearly indicate, via low-level error requirements, what order is appropriate for reduction. This advantage is especially evident for interconnections of many subsystems, where it becomes increasingly difficult to select an appropriate reduced order per subsystem.

\subsection{Practical aspects/considerations for applying robust abstracted reduction} \label{ssec:robabsred_practical}
The robust abstracted reduction frameworks in Algorithms \ref{alg:robabsred_env} and \ref{alg:robabsred_ssys} require an initial selection of $G_u$, $G_y$, $\bm{\Vc}$ and $\bm{W}$ (specifically, $\bm{\Vc}_E,\ \bm{W}_E$ for \autoref{alg:robabsred_env} and $\bm{\Vc}_\sys,\ \bm{W}_\sys$ for \autoref{alg:robabsred_ssys}). Their selection influences the resulting reduced-order model and the conservatism of the results obtained using that selection. Below, we outline some practical considerations and implications of these selections.

The main observation is that robust abstracted reduction uses \Hnrm{\infty}-bounds, which can be conservative if the considered transfer function magnitude shows large fluctuations in magnitude. For instance, if $\eb_E^1 = \|(W^{1}_E)^{-1}\sees{1}{22}(\Vc^{1}_E)^{-1}\|_\infty \gg (W^{1}_E)^{-1}\sees{1}{22}(\Vc^{1}_E)^{-1}$ for most frequencies, this can lead to excessive conservatism. To avoid this, the weighted errors, as in \autoref{eq:eb_def}, should be as uniform (over the relevant frequency range) as possible. This can be achieved by selecting weighting matrices $\bm{\Vc}$ and $\bm{W}$ that match the expected error $\bm{\Lambda}$ ($\bm{\Lambda}_E$ or $\bm{\Lambda}_\sys$), but this may be a difficult task as $\bm{\Lambda}$ is usually unknown. 

An alternative and more practical approach involves using weighted reduction methods that emphasize the accuracy in frequency ranges where the requirements, defined by $\bm{\Vc}$ and $\bm{W}$, are strict. This effectively shapes the error $\bm{\Lambda}$ such that $\bm{W}^{-1}\bm{\Lambda}\bm{\Vc}^{-1}$ is more uniform. Specifically, in the case of subsystem abstraction, $\|(W^{j}_A)^{-1}\sea{j}(\Vc^{j}_A)^{-1}\|_\infty$ can be minimized by using $(\Vc^{j}_A)^{-1}$ and $(W^{j}_A)^{-1}$ as input- and output-weighting, respectively. 

As opposed to the specification for subsystem abstraction, the accuracy specifications for environment abstraction and subsystem reduction of \autoref{eq:eb_def} only reflect the 22-partitions of $\senv{j}$ and $\llft(\sfnva{j},\ssys{j})$. However, the accuracy of other partitions also influences the overall accuracy of $\llft(S,\sysr_B)$ and should not be overlooked. For environment abstraction, it is thus recommended to use $G_{i}^j$ and $G_{o}^j$ for the input and output weights, respectively, as defined by
\begin{equation} \label{eq:abs_IO_weights}
    G_{i}^j = \diag(1,(\Vc^{j}_E)^{-1}),\quad G_{o}^j = \diag(1,(W^{j}_E)^{-1}).
\end{equation}
In the case of subsystem reduction, input and output weights are already embedded in $\sfnva{j}$ as $G_{u}^j$ and $G_{y}^j$, respectively. This ensures that a structure-preserving reduction of $\llft(\sfnva{j},\ssys{j})$ approximates $(W^{j}_F)^{-1}\llft(\sfnva{j},\ssys{j})_{22}(\Vc^{j}_F)^{-1}$ by setting
\begin{equation} \label{eq:red_IO_weights}
    G_{u}^j = (\Vc^{j}_F)^{-1},\quad G_{y}^j = (W^{j}_F)^{-1}.
\end{equation} 

\begin{sloppypar}
Although these weighted reduction methods with weights as in \autoref{eq:abs_IO_weights} and \autoref{eq:red_IO_weights}, compensate for the selected $\bm{\Vc}$ and $\bm{W}$ such that $\bm{W}^{-1}\bm{\Lambda}\bm{\Vc}^{-1}$ is mostly uniform, the choice of $\bm{\Vc}$ and $\bm{W}$ is not without consequence. Therefore, we present some considerations in the actual selection of $\bm{\Vc}$ and $\bm{W}$: 
\end{sloppypar}
\begin{enumerate}
    \item All weighting functions should be bistable for the use of \autoref{thm:req_validation}.
    \item The magnitudes of $W^{j}_E$, $\Vc^{j}_E$, $W^{j}_F$ and $\Vc^{j}_F$ should be selected appropriately, balancing the accuracy of the various partitions of $\senv{j}$ and of $\llft(\sfnva{j},\ssys{j})$.
    \item The weights should be low-order transfer functions to avoid unnecessary complication of the weighted reduction.
\end{enumerate}

Regarding point 2, note that if the reduction is weighted as in \autoref{eq:abs_IO_weights} and \autoref{eq:red_IO_weights} and the magnitudes of $W^{j}_E$, $\Vc^{j}_E$, $W^{j}_F$ and $\Vc^{j}_F$ are low, the 22-partitions are typically approximated more accurately. However, this high accuracy of the 22-partitions may come at the expense of lower accuracy in the other partitions, such that low-magnitude weights can result in $\llft(\senva{j},\ssys{j})\not\approx\llft(\senv{j},\ssys{j})$ and $\llft(\senva{j},\ssysr{j})\not\approx\llft(\senva{j},\ssys{j})$. Low-magnitude weighting therefore typically allows for further reduction of $E$ and $\sys$ while simultaneously reducing the overall accuracy of $\llft(S,\sysr_B)$. Conversely, high-magnitude weighting results in a more accurate $\llft(S,\sysr_B)$ but is more conservative, resulting in less reduction of $\env$ and $\sys$. A balance should thus be found.

Regarding point 3, $\bm{\Vc}$ and $\bm{W}$ should be selected such that the weighted reduction method can easily reduce the considered model, minimizing the weighted \Hnrm{\infty}-bound. For example, in case the dynamical weighting functions require an especially strict accuracy in hard-to-approximate (e.g., near sharp resonances) frequency ranges or have sharp peaks themselves, the \Hnrm{\infty}-bounds may be more conservative and the model may not be reduced as far as desired. We therefore suggest to use simple high-pass or low-pass filters that match the general trend of the model's transfer function magnitude, allowing the weighted error to remain relatively uniform over the relevant frequency range.

\section{Abstracted reduction of an interconnected, structural-dynamics model}\label{sec:case_study_red}
To evaluate the performance of the two abstracted reduction approaches and the framework for error analysis, we will reduce an interconnected, structural-dynamics model and compare the approaches to the open-loop reduction approach of \cite{Janssen2023ModularApproach}. Besides reducing the subsystem models independently, the approach of \cite{Janssen2023ModularApproach} uses similar techniques from robust performance analysis to guarantee stability and accuracy of the reduced, interconnected system model, such that the influence of using abstracted reduction can be properly compared. Other methods to reduce the subsystem models within an interconnection typically do not provide accuracy and stability guarantees and thus do not offer such a fair comparison.

The model we evaluate is based on equipment in the lithography industry, specifically a wafer stage with a long-stroke and short-stroke motion stage and an illuminator above, as treated in \cite{Butler2013PositionEquipment}. A more detailed description of the model will be given in \autoref{ssec:waferstage_intro}. To compare the reduction methods, all subsystem models will be reduced such that the reduced interconnected model meets a prescribed accuracy specification. In \autoref{ssec:robabsred_eval}, the different reduction approaches are then assessed and compared on the basis of the minimum order that is required to meet this specification and on the resulting accuracy of the reduced interconnected model. In \autoref{ssec:conservatism_eval}, we will further evaluate the sources of conservatism within the abstracted reduction framework, after which key insights are summarized in \autoref{ssec:numeval_insights}.

\subsection{A 2D dynamic wafer stage model} \label{ssec:waferstage_intro}
\begin{figure}
    \centering    
    \includegraphics[width=0.8\linewidth]{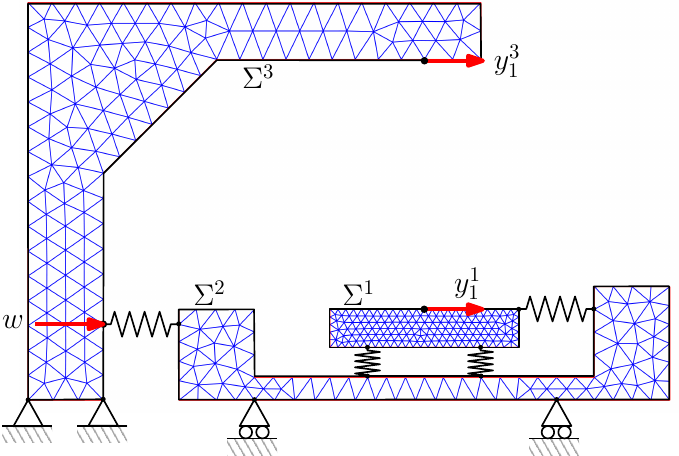}
    \caption{Schematic illustration of the 2D wafer stage.}
    \label{fig:2dwaferstage}
\end{figure}

The 2D system, as schematically shown in a vertical plane in \autoref{fig:2dwaferstage}, consists of three components: the wafer ($\ssys{1}$) in the middle, the stage ($\ssys{2}$) on the bottom right and the illuminator frame ($\ssys{3}$, illuminator for short) on the left. The illuminator is fixed at its two bottom corners in both the horizontal and vertical direction, whereas the stage is fixed at two points in only the vertical direction, such that it can translate freely in horizontal direction. The wafer is not connected to the fixed world.

Each component is modelled using finite element meshing with quadratic, 3-node triangular plane-stress elements with two translational degrees of freedom per node, as visualized in \autoref{fig:2dwaferstage}. To simplify computation, an initial reduction is performed using the Craig-Bampton reduction method \cite{Craig1968CouplingAnalyses}, such that each component model consists of 100 states, see also \autoref{tab:subsys_counts}. These component models with 100 states  are considered to be the unreduced component models as a starting point before applying the abstracted reduction framework. Furthermore, gravity is neglected because it is not important for demonstrating the framework.

\begin{table}[]
\caption{Number of states $n_j$, inputs $m_j$ and outputs $p_j$ per subsystem.}
\label{tab:subsys_counts}
\centering
\begin{tabular}{llcccc}  \toprule
Symbol      &  Name                  & $n_j$ & $m_j$ & $p_j$  \\ \midrule
$\ssys{1}$  &  Wafer                 & 100   & 3     & 4      \\
$\ssys{2}$  & Stage                  & 100   & 4     & 4      \\
$\ssys{3}$  & Illuminator            & 100   & 1     & 2      \\ \hline
$\sys_C$    & Interconnected system  & 300   & 1     & 1      \\ \bottomrule
\end{tabular}
\end{table}

The components are interconnected by translational springs, see \autoref{fig:2dwaferstage}. The vertical springs, with a stiffness of $k_c = 1\times 10^{11}\,$N/m, model the vertical stiffness of the suspension between the wafer and the stage. We assume that both the long-stroke actuation, between the stage and the illuminator, and the short-stroke actuation, between the stage and the wafer, can be approximated by translational horizontal springs with stiffnesses $k_l = 1\times 10^{11}\,$N/m and $k_s = 1\times 10^{13}\,$N/m, respectively.

We consider a disturbance attenuation problem of this wafer stage model, in which the horizontal external input force $w$ [N] is a disturbance at the illuminator (see \autoref{fig:2dwaferstage}) and the external output $z$ [m] is equal to the horizontal misalignment (overlay) between wafer and illuminator $y^{3}_1-y^{1}_1$. This results in the static interconnection
\begin{equation}
     S =   \mbox{\scriptsize$ \left[
    \begin{array}{c|cccccccccc}
        \!0 \!\!&\!\! -1\!\!&\!\! 0   \!\!&\!\! 0  \!\!&\!\! 0  \!\!&\!\! 0  \!\!&\!\! 0  \!\!&\!\! 0  \!\!&\!\! 0  \!\!&\!\! 1 \!\!&\!\! 0 \!\!\!\!\\ \hline
        \!0 \!\!&\!\! 0 \!\!&\!\!\!-k_s \!\!&\!\! 0  \!\!&\!\! 0  \!\!&\!\! k_s\!\!&\!\! 0  \!\!&\!\! 0  \!\!&\!\! 0  \!\!&\!\! 0 \!\!&\!\! 0  \!\!\!\! \\
        \!0 \!\!&\!\! 0 \!\!&\!\! 0   \!\!&\!\!\!-k_c\!\!&\!\! 0  \!\!&\!\! 0  \!\!&\!\! 0  \!\!&\!\! 0  \!\!&\!\! 0  \!\!&\!\! 0 \!\!&\!\! 0 \!\!\!\!\\
        \!0 \!\!&\!\! 0 \!\!&\!\! 0   \!\!&\!\! 0  \!\!&\!\!\!-k_c\!\!&\!\! 0  \!\!&\!\! 0  \!\!&\!\! 0  \!\!&\!\! 0  \!\!&\!\! 0 \!\!&\!\! 0 \!\!\!\!\\
        \!0 \!\!&\!\! 0 \!\!&\!\! k_s \!\!&\!\! 0  \!\!&\!\! 0  \!\!&\!\!\!-k_s\!\!&\!\! 0  \!\!&\!\! 0  \!\!&\!\! 0  \!\!&\!\! 0 \!\!&\!\! 0 \!\!\!\!\\
        \!0 \!\!&\!\! 0 \!\!&\!\! 0   \!\!&\!\! 0  \!\!&\!\! 0  \!\!&\!\! 0  \!\!&\!\!\!-k_l\!\!&\!\! 0  \!\!&\!\! 0  \!\!&\!\! 0 \!\!&\!\! k_l \!\!\!\!\\
        \!0 \!\!&\!\! 0 \!\!&\!\! 0   \!\!&\!\!k_c \!\!&\!\! 0  \!\!&\!\! 0  \!\!&\!\! 0  \!\!&\!\!\!-k_c\!\!&\!\! 0  \!\!&\!\! 0 \!\!&\!\! 0 \!\!\!\!\\
        \!0 \!\!&\!\! 0 \!\!&\!\! 0   \!\!&\!\! 0  \!\!&\!\! k_c\!\!&\!\! 0  \!\!&\!\! 0  \!\!&\!\! 0  \!\!&\!\!\!-k_c\!\!&\!\! 0 \!\!&\!\! 0\!\!\!\! \\
        \!1 \!\!&\!\! 0 \!\!&\!\! 0   \!\!&\!\! 0  \!\!&\!\! 0  \!\!&\!\! 0  \!\!&\!\!k_l \!\!&\!\! 0  \!\!&\!\! 0  \!\!&\!\! 0 \!\!&\!\! \!-k_l\!\!\!\!
    \end{array} \right],$ }
\end{equation}
such that the full interconnected model $\llft(S,\sys_B)$, where $\sys_B = \diag(\ssys{1},\ssys{2},\ssys{3})$, is a SISO system of 300 states.

\subsection{Robust abstracted reduction}\label{ssec:robabsred_eval}
In this section, we will reduce each subsystem model such that the reduced, interconnected model satisfies a prescribed accuracy specification. We will compare robust abstracted reduction by environment abstraction, henceforth denoted RAR-$E$, and by subsystem abstraction (RAR-$\sys$) to one another and to the robust subsystem reduction method (RSS) of \cite{Janssen2023ModularApproach}. 

Let us first prescribe the interconnected accuracy specification 
$\uec$, as in \autoref{eq:uec_def}, by selecting $W_C = 1$ and 
\begin{equation}\label{eq:imposed_ecb}
V_C^{-1} = \tfrac{1}{4}\,\llft\big(S,\sys_B\big) + 5\times 10^{-12}.
\end{equation}

\begin{figure}
    \centering
    \includegraphics[width = 0.9\linewidth]{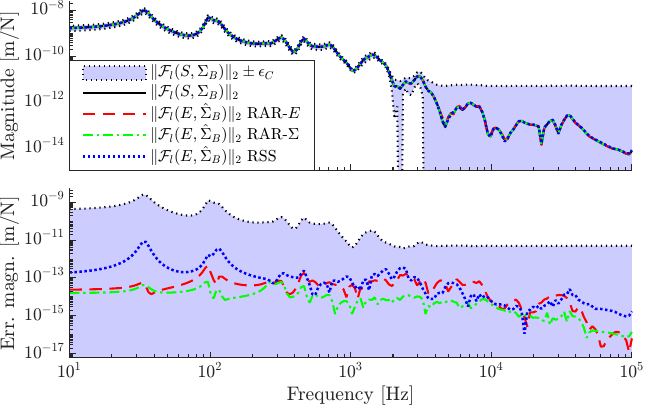}
    \caption{Top plot: Magnitudes of the frequency responses of the interconnected model $\llft(S,\sys_B)$ and the approximate models following from RAR-$E$, RAR-$\sys$ and RSS \cite{Janssen2023ModularApproach}. Bottom plot: the error magnitudes for the various approximate models. In both plots, the high-level accuracy specification $\epsilon_C$ is indicated by the blue area.
    }
    \label{fig:FRF-ss&env_DC_BT}
\end{figure}

To interpret this specification more easily, we define $\epsilon_C(\omega) = \|V_C^{-1}(i\omega)\|$ for all $\omega \in \R$. This makes $\ec\in\uec$ equivalent to $\|\ec(i\omega)\| \leq \epsilon_C(\omega)$ for all $\omega \in \R$. This accuracy specification can be visualized in the frequency domain as a band around the magnitude of the frequency response of $\llft(S,\sys_B)$, as shown by the blue area in \autoref{fig:FRF-ss&env_DC_BT}. For this specification to be satisfied the frequency response of $\llft(S,\sysr_B)$ should be fully contained within in this blue area. Note that the second (constant) term on the right hand side of \autoref{eq:imposed_ecb} causes a very lenient relative accuracy requirement on the reduced interconnected system model at frequencies above 4000 Hz. This is also reflected by the wide blue band in the top plot of \autoref{fig:FRF-ss&env_DC_BT}.

\begin{figure}
    \subfloat[\small]{%
      \includegraphics[width=.48\linewidth]{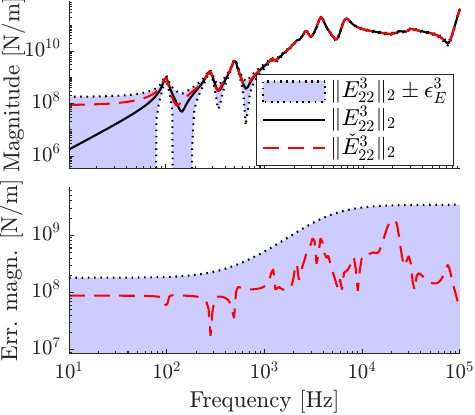} \label{fig:Henv_DE3_BT}
    }\hfill
    \subfloat[\small]{%
      \includegraphics[width=.48\linewidth]{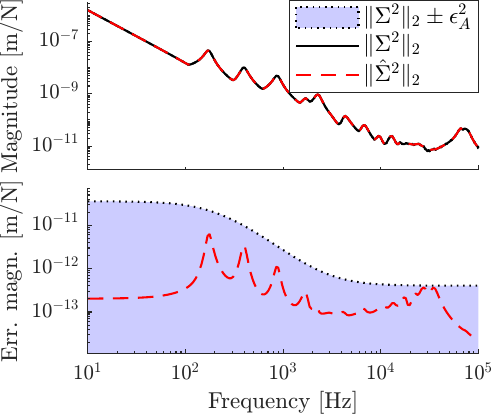} \label{fig:HSS_DS2_BT}
    } \\
    \subfloat[\small]{%
      \includegraphics[width=.48\linewidth]{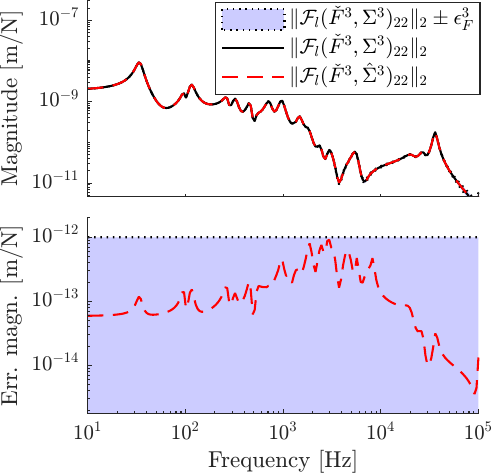} \label{fig:Henv_DF3_BT}
    }\hfill
    \subfloat[\small]{%
      \includegraphics[width=.48\linewidth]{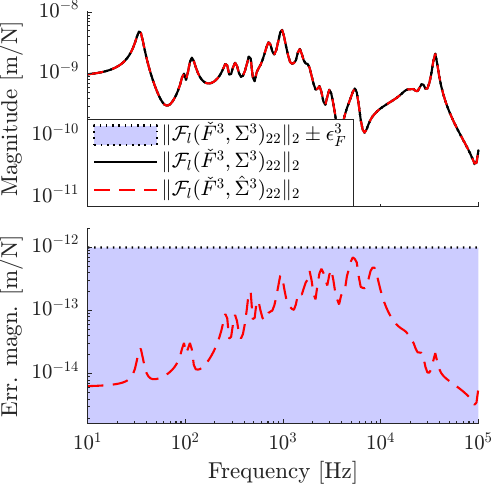} \label{fig:Hss_DF3_BT}
    }
    \caption{Spectral norms of the frequency response matrices of the full-order models (\textcolor{black}{\textbf{---}}), reduced-order models (\textcolor{red}{\textbf{--\,--}}) and their allow error bounds (\,\errboundbox\,), following from \autoref{eq:eb_VW_def}, for (a) the environment abstraction of $\senv{3}_{22}$, (b) the subsystem abstraction of $\ssys{2}$ and the structure-preserving reduction of $\llft(\sfnva{3},\ssys{3})$ for (c) environment abstraction and (d) subsystem abstraction.}
    \label{fig:FRF-Hinf_BT}
\end{figure}

As a first step, we translate the high-level accuracy specification $\uec$ to low-level accuracy specifications. For RAR-$E$, RAR-$\sys$, this is realized by solving the optimization problem of \autoref{th:opt_alg_TD_infty}, whereas the optimization problem for RSS is given in \cite{Janssen2023ModularApproach}. All optimization problems are solved using MOSEK \cite{MOSEKApS2024MOSEK10.2.} in combination with Yalmip \cite{Lofberg2004YALMIP:MATLAB} and MATLAB \cite{MATLAB}. The weighting functions $\bm{\Vc}$ and $\bm{W}$ are prescribed as diagonal transfer function matrices of first-order high-pass ($\Vc^{j}_E$ and $W^{j}_E$) or low-pass ($\Vc^{j}_A$, $W^{j}_A$, $\Vc^{j}_F$ and $W^{j}_F$) filters to match frequency response magnitude of the corresponding systems (see also \autoref{ssec:robabsred_practical}). Some examples of the resulting low-level accuracy specifications are visualized by the blue area's in \autoref{fig:FRF-Hinf_BT}, where each bound is given as
\begin{equation} \label{eq:eb_VW_def}
\begin{gathered}
    \epsilon^{j}_X = \|J\|^{-1}\, W_X^{j} \,J\,V_X^{j}, \\ \text{for } j = \{1,2,3\}, \ X = \{A,E,F\},
\end{gathered}
\end{equation}
where $J$ is a matrix of appropriate dimensions filled with ones.

Subsequently, either the environment models $\senv{j}$ (with RAR-$E$) or subsystem models $\ssys{j}$ (with RAR-$\sys$) are abstracted to an order that is as low as possible, while still satisfying the accuracy specifications of $\suees{j}{22}$ and $\suea{j}$, respectively. To avoid conservatism in the abstraction process, frequency-weighted balanced truncation (FWBT) \cite{Antoulas2005ApproximationSystems,Enns1984ModelGeneralization} is used for the abstraction, minimizing $\|(W^{j}_A)^{-1}\sea{j}(V^{j}_A)^{-1}\|_\infty$ and  $\|G_{o}^j\,\see{j}\,G_{i}^j\|_\infty$ for RAR-$\sys$ and RAR-$E$, respectively, with $G_{o}^j$ and $G_{i}^j$ as in \autoref{eq:abs_IO_weights}.

The resulting abstract environments $\senva{j}$ are then augmented with $G_u^{j}$ and $G_y^{j}$ as in \autoref{eq:red_IO_weights} to obtain $\sfnva{j}$, as shown in \autoref{eq:F_defs}. The resulting interconnected model $\llft(\sfnva{j},\ssys{j})$ is then reduced to $\llft(\sfnva{j},\ssysr{j})$ using interconnected systems balanced truncation (ISBT) \cite{Sandberg2009,Vandendorpe2008ModelSystems}, effectively reducing each subsystem model $\ssys{j}$.


In \autoref{tab:red_orders}, the reduced model orders resulting from RAR-$E$ and RAR-$\sys$ are compared to the orders resulting from robust subsystem reduction (RSS), as presented in \cite{Janssen2023ModularApproach}. All three methods guarantee that the reduced interconnected system is stable and the introduced error $\ec$ satisfies the prescribed accuracy specification of \autoref{eq:imposed_ecb}. However, RAR-$E$ clearly results in the largest reduction, whereas RSS and and RAR-$\sys$ result in reduced models of higher order.

Although RAR-$E$ reduces further than RSS, RAR-$E$ requires more computational resources for the structure-preserving reduction of $\llft(\sfnva{j},\ssys{j})$ than RSS requires for the subsystem reduction of $\ssys{j}$. RAR-$\sys$ is even more computationally costly than RAR-$E$, because the abstracted environments $\senva{j}$ are of higher order, as shown in \autoref{tab:red_orders}. In this example, the environments are not abstracted to a much lower order due to the relatively small order and small number of subsystems, making much of the environments' internal dynamics relevant to the interconnected system. Typically, environments of higher orders can be reduced significantly further, such that an abstracted environment might be of a lower order than the corresponding subsystem. Consequently, for (very) high-order interconnected system models, the structure-preserving reduction using RAR-$E$ and RAR-$\sys$ is likely to have a similar cost to the subsystem reduction of RSS (see also \cite[Section~III-B]{Poort2024AbstractedReduction}).

\begin{table}[]
\centering
\caption{Full orders ($n_j$) and the reduced subsystem, environment and interconnected model orders resulting from the three robust reduction approaches RAR-$E$, RAR-$\Sigma$ and RSS.}
\label{tab:red_orders}
\begin{tabular}{c|c|ccc} \toprule
System                & $n_j$ & RAR-$E$ & RAR-$\Sigma$ & RSS \\ \midrule
$\ssysr{1}$           & 100   & 40      & 64  & 50         \\
$\ssysr{2}$           & 100   & 56      & 74  & 68         \\
$\ssysr{3}$           & 100   & 32      & 40  & 38         \\
$\senva{1}$           & 200   & 56      & 90  & -         \\
$\senva{2}$           & 200   & 70      & 94  & -          \\
$\senva{3}$           & 200   & 26      & 116 & -           \\ \hline
$\llft(S,\sysr_B)$    & 300   & 128     & 178 & 156       \\ \bottomrule
\end{tabular}
\end{table}

While all three reduced models satisfy the prescribed interconnected accuracy specification of \autoref{eq:imposed_ecb}, the three reduction methods differ in their actual accuracy, as shown in \autoref{fig:FRF-ss&env_DC_BT}. All three models are highly accurate with error magnitudes one to four orders lower than the prescribed bound. The most accurate reduced-order model is attained by RAR-$\sys$, but its order is also the highest, as shown in \autoref{tab:red_orders}. Interestingly, RAR-$E$ results in a higher accuracy than RSS, even though the interconnected model is reduced by an additional 28 states.

Summarizing, RAR-$E$ reduces the interconnected system the most while being more accurate than the existing approach from literature, RSS \cite{Janssen2023ModularApproach}, and thus seems the best choice for this benchmark. RAR-$\sys$ does not reduce as far, but results in the most accurate model and is thus more conservative (in relation to the error specification).  However, RAR-$\sys$ provides a higher level of modularity, as previously discussed in \autoref{sec:absred}. Therefore, if a high level of modularity is required, such as in the first stages of the design of a new interconnected system, and the system is very strongly connected, such that RSS is insufficient, RAR-$\sys$ may very well have added value. 

As all three reduction methods display conservatism, producing errors which are more than a factor 10-100 times lower than the imposed error bound, the origin of this conservatism is analyzed next.

\subsection{Sources of conservatism} \label{ssec:conservatism_eval}
The conservatism of the reduced models in \autoref{fig:FRF-ss&env_DC_BT} can be explained by two different causes: 1) the conservatism of the translation of the high-level bound \autoref{eq:imposed_ecb} into the low-level bounds and 2) the conservatism of the actual reduction and abstraction steps. The latter can be observed rather straightforwardly in \autoref{fig:FRF-Hinf_BT}, where the frequency response of several reduced-order models is visualized as the red dashed lines. Clearly, the imposed low-level accuracy specifications are not fully exploited by the balanced reduction method, resulting in a reduction error that is significantly lower than the actual bound at most frequencies. These low-level sources of conservatism obviously also contribute to the eventual conservatism of the reduced interconnected model. 

As RAR-$E$ and RAR-$\sys$ include an additional abstraction step, twice as many low-level accuracy specifications have to be satisfied when compared to RSS, typically resulting in an increased conservativeness. However, the main premise of abstracted reduction is that the accuracy of the reduced interconnected model is less sensitive to errors resulting from abstraction and structure-preserving reduction, therefore leading to a more accurate interconnected model.

\begin{figure}
    \subfloat[\small]{%
      \includegraphics[width=.48\linewidth]{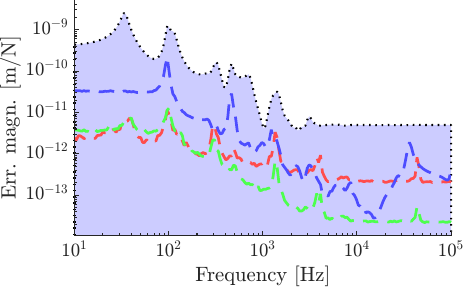} \label{fig:FRF-DC_Hmax}
    }\hfill
    \subfloat[\small]{%
      \includegraphics[width=.48\linewidth]{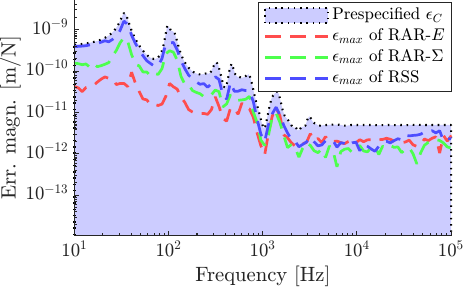} \label{fig:FRF-DC_max}
    } 
    \caption{Frequency response magnitude of the high-level accuracy specification $\epsilon_C$ and the maximum attainable errors, utilizing the full low-level accuracy specifications $\suees{j}{22}$, $\suea{j}$ and $\sueft{j}$, using (a) first-order weighting functions and (b) optimized, diagonal weighting functions.}
    \label{fig:FRF-DCmax_both}
\end{figure}

To test the first source of conservatism, i.e., the one introduced by translating high-level specifications to low-level specifications, we will artificially construct random, approximate dynamical systems $\senva{j}$, $\ssysa{j}$ and $\ssysr{j}$, for which the corresponding errors $\sees{j}{22}$, $\sea{j}$ and $\seft{j}$ always meet the specifications $\suees{j}{22}$, $\suea{j}$ and $\sueft{j}$. Specifically, we generate random, complex matrices $R_E,\ R_A,\ R_F$ of spectral norm at most $1$ and, for a discrete set of one hundred, logarithmatically-spaced frequencies $\bbOmega \coloneqq \{\omega_1,\dots,\omega_{100}\}$, we determine the approximate models as
\begin{equation}
    \begin{aligned}
        \senva{j}_{22}(i\omega_q) &= \senv{j}_{22}(i\omega_q) + \sW{j}_E(i\omega_q)\, R_E\, \sV{j}_E(i\omega_q) ,\\
        \ssysa{j}(i\omega_q) &= \ssys{j}(i\omega_q) +  \sW{j}_A(i\omega_q)\, R_A\, \sV{j}_A(i\omega_q) ,\\
        \seft{j}(i\omega_q) &= \sW{j}_F(i\omega_q)\, R_F\, \sV{j}_F(i\omega_q),
    \end{aligned}
\end{equation}
using first-order, high-pass and low-pass filters as weighting functions (as before). The approximate subsystem's frequency response $\ssysr{j}(i\omega_q)$ is subsequently determined using \autoref{eq:Eclem_sysrb}, after which the approximate interconnected frequency response is found by interconnecting all $k$ matrices $\ssysr{j}(i\omega_q)$ with $S$ as $\llft\big(S,\sysr_B(i\omega_q)\big)$. In this way, one thousand different interconnected ``models'' were generated, for different $R_E,\ R_A,\ R_F$, and the maximum absolute error per frequency point $\omega_q$ is visualized in \autoref{fig:FRF-DC_Hmax}, which shows varying conservatism with respect to the prescribed error bound of $\epsilon_{C}$. 

This conservatism, although limited, has two causes:
\begin{enumerate}
    \item The low-level error bounds, as visualized for example in \autoref{fig:FRF-Hinf_BT}, are weighted $H_\infty$-bounds, where the weighting is selected beforehand. Such $H_\infty$-bounds show quite some conservatism with respect to the actual maximal error per frequency point.
    \item The framework itself is slightly conservative, as an upper bound on the structured singular value $\mu$ is used instead of its actual value (see also \cite{Janssen2024ModularPerspective}).
\end{enumerate}

To evaluate the influence of weighted $H_\infty$-bounds, an alternative optimization algorithm is attempted, which optimizes \emph{diagonal} $\sV{j}_X(i\omega_q)$ and $\sW{j}_X(i\omega_q)$ directly, for $X \in \{A,E,F\},\ j = 1,2,3, \omega_q \in \bbOmega$. This approach does not guarantee stability but allows to find weighting matrices per frequency point, resulting in much less conservative weighting matrices, see \cite[Theorem~1]{Janssen2023ModularApproach} and \cite[Theorem~4]{Poort2024AbstractedReduction} for details. The resulting maximum errors using these new weighting matrices are visualized in \autoref{fig:FRF-DC_max}, and are significantly less conservative.

The decrease in conservatism when comparing \autoref{fig:FRF-DC_Hmax} to \autoref{fig:FRF-DC_max} indicates that the choices of weighting functions largely determine the conservatism of the method. The remaining conservatism can be attributed to the weighting matrices being diagonal and the conservativeness in the upper bound of the structured singular value $\mu$.

Comparing the three reduction methods, \autoref{fig:FRF-DCmax_both} shows that the framework of RSS is the least conservative. Although RAR-$\sys$ seems to be the most conservative in \autoref{fig:FRF-DC_Hmax}, its conservatism is significantly smaller in \autoref{fig:FRF-DC_max}. This may indicate that the weights $\sW{j}_A$ and $\sV{j}_A$ were less suitably chosen than the other weights, resulting in the large conservatism in \autoref{fig:FRF-DC_Hmax} and perhaps leading to the relatively high reduced order in \autoref{tab:red_orders}.

\subsection{Key insights from the numerical evaluation} \label{ssec:numeval_insights}
For the benchmark introduced in \autoref{ssec:waferstage_intro}, abstracted reduction using environment abstraction (RAR-$E$) outperforms the existing robust subsystem reduction approach (RSS) of \cite{Janssen2023ModularApproach} in both the achieved reduction and resulting accuracy. These improvements are attributed to the consideration of (an abstraction of) the environments of the subsystems.

The abstracted reduction approach using subsystem abstraction (RAR-$\sys$) produces a highly accurate model but achieves less reduction compared to RSS and RAR-$E$. While including an abstracted environment enhances the accuracy, in this specific case, RAR-$\sys$ introduces excessive conservatism, making it difficult to justify its use.

Further analysis of this conservatism indicates that the selected weighting functions $\bm{\Vc}$ and $\bm{W}$ significantly impact the conservatism of the approach.

%
%
\section{Conclusions} \label{sec:con}
We have introduced two variants of the framework of abstracted reduction to improve the tractability of structure-preserving reduction of interconnected subsystems. In this framework, the structure-preserving reduction method is not applied to the full interconnection of all subsystem models, but to each subsystem model connected to a low-order abstraction of its environment. The two variants differ in the construction of these low-order environment abstractions, balancing accuracy of this abstraction with the modularity of the approach. By thus having reduced the order of the abstracted, interconnected model, the computational cost of reducing the subsystems is significantly reduced. Leveraging techniques from robust performance analysis, we introduced a second approach to also automatically determine appropriate abstraction and reduction orders that guarantee 1) the reduced interconnected system model to be stable and 2) a prescribed $\Hnrm{\infty}$-accuracy specification of the reduced interconnected system model.

In a numerical study on an industrial benchmark system, we evaluated both variants of robust abstracted reduction approaches against the existing robust subsystem reduction method described by \cite{Janssen2023ModularApproach}. Specifically, we applied these reduction techniques to a 2D wafer stage structural-dynamics model. Our findings indicate that the structure-preserving reduction method utilizing an abstracted environment outperforms the approach by \cite{Janssen2023ModularApproach} by achieving both a higher accuracy and lower order of the reduced interconnected model.


%
%
\section*{Acknowledgements}
The authors would like to thank Dr. Victor Dolk and Thijs Verhees, MSc, for valuable discussions.
%
%
\printcredits
%
%
\section*{Declaration of competing interest} 
The authors have no competing interests to declare that are relevant to the content of this article.

\bibliographystyle{IEEEtran}
\bibliography{references}

\end{document}

%% file: cmd_def.tex
\usepackage{amsmath}
\usepackage{amssymb}
\usepackage{mathtools}
\usepackage{bm}
\usepackage{booktabs}
\usepackage{lipsum}
\usepackage{xcolor}
\usepackage{doi}
\usepackage[numbers]{natbib}
\usepackage{ifmtarg}
\usepackage{tikz}
\usepackage[shortlabels]{enumitem}

\definecolor{ML_c1}{rgb}{0, 0.447, 0.741}
\definecolor{ML_c2}{rgb}{0.85, 0.325, 0.098}
\definecolor{ML_c3}{rgb}{0.929, 0.694, 0.125}
\definecolor{ML_c4}{rgb}{0.494, 0.184, 0.556}

\usepackage{amsthm}
\usepackage{aliascnt}                       
\theoremstyle{remark}
\newtheorem{thm}{Theorem}
\newtheorem{lem}{Lemma}

\newtheorem{defn}{Definition}
\newtheorem{rem}{Remark} 
\newtheorem{ass}{Assumption}

\newtheorem{alg}{Algorithm}
\long\def\@makealgocaption#1#2{\vskip 2ex \small
  \hbox to \hsize{\parbox[t]{\hsize}{{\bfseries #1.} #2}}}
\newcounter{algorithm}
\def\thealgorithm{\@arabic\c@algorithm}
\def\fps@algorithm{tbp}
\def\ftype@algorithm{4}
\def\ext@algorithm{lof}
\def\fnum@algorithm{Algorithm \thealgorithm}

\def\equationautorefname~#1\null{%
  (#1)\null
}

\usepackage[outline]{contour}
\newcommand*\errboundbox{\raisebox{-.2mm}{\includegraphics[scale =0.28]{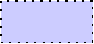}}}

\newcommand*\bbOmega{\includegraphics{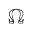}}
\newcommand*\bbLambda{\includegraphics[scale =0.7]{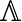}}

\newcommand\shdots{\makebox[1em][c]{.\hfil.\hfil.}}

\newcommand{\beq}{\begin{equation}}
\newcommand{\eeq}{\end{equation}}
\newcommand{\beqa}{\begin{equation} \begin{aligned}}
\newcommand{\eeqa}{\end{aligned} \end{equation}}

\DeclareMathOperator\diag{diag}

\newcommand{\R}{\mathbb{R}}
\newcommand{\C}{\mathbb{C}}

\newcommand{\Hnrm}[1]{{\ensuremath{\mathcal{H}_{#1}}}}
\newcommand{\RH}{\mathcal{R}\Hnrm{\infty}}

\newcommand{\llft}{\mathcal{F}_l}       
\newcommand{\ulft}{\mathcal{F}_u}       

\newcommand{\sys}{\Sigma}           	
\newcommand{\sysr}{\hat{\sys}}          
\newcommand{\sysa}{\check{\sys}}        
\newcommand{\ssys}[1]{\sys^{#1}}        
\newcommand{\ssysb}[1]{\bar{\sys}^{#1}}    
\newcommand{\ssysr}[1]{\sysr^{#1}}      
\newcommand{\ssysa}[1]{\sysa^{#1}}      
\newcommand{\ssysba}[1]{\check{\bar{\sys}}^{#1}}    

\renewcommand{\S}{S}
\newcommand{\Sb}[1]{\bar{\S}^{#1}}

\newcommand{\env}{E}           	        
\newcommand{\enva}{\check{\env}}        
\newcommand{\envb}{\bar{\env}}          
\newcommand{\senv}[1]{\env^{#1}}      
\newcommand{\senva}[1]{\enva^{#1}}    
\newcommand{\senvb}[1]{\envb^{#1}}    

\newcommand{\fnv}{F}           	        
\newcommand{\fnva}{\check{\fnv}}        
\newcommand{\sfnva}[1]{\fnva^{#1}}    

\newcommand{\efs}[1]{\Lambda_{F,#1}}         
             
\newcommand{\ec}{\Lambda_C}     
\newcommand{\see}[1]{\Lambda_E^{#1}} 
\newcommand{\sea}[1]{\Lambda_A^{#1}} 
\newcommand{\seab}[1]{\bar{\Lambda}_A^{#1}} 
\newcommand{\sees}[2]{\Lambda_{E,#2}^{#1}}              
\newcommand{\sef}[1]{\Lambda_F^{#1}}        
\newcommand{\sefs}[2]{\Lambda_{F,#2}^{#1}}        
\newcommand{\seft}[1]{\tilde{\Lambda}_F^{#1}}     

\newcommand{\ue}{\bbLambda}

\newcommand{\uec}{\ue_C}    
 
\newcommand{\suea}[1]{\ue_A^{#1}} 
 
\newcommand{\suees}[2]{\ue_{E,#2}^{#1}}

\newcommand{\sueft}[1]{\tilde{\ue}{\vphantom{\ue}}^{#1}_F}   

\newcommand{\eb}{\varepsilon}
\newcommand{\ebb}{\bar{\varepsilon}}

\newcommand{\sebe}[1]{\eb_E^{#1}}    
\newcommand{\seba}[1]{\eb_A^{#1}}    
\newcommand{\sebba}[1]{\ebb_A^{#1}}                
\newcommand{\sebf}[1]{\eb_F^{#1}}                

\newcommand{\sV}[1]{V^{#1}}
\newcommand{\sW}[1]{W^{#1}}
\newcommand{\sVb}[1]{\bar{V}^{#1}}
\newcommand{\sWb}[1]{\bar{W}^{#1}}
\newcommand{\Vc}{\mathcal{V}}

\newcommand{\sVc}[1]{\Vc^{#1}}

\newcommand{\sVbc}[1]{\bar{\Vc}^{#1}}